\documentclass[%
reprint,
superscriptaddress, 
amsmath, amssymb, 
aps, 
pra,
longbibliography]{revtex4-1}

\usepackage[utf8]{inputenc}
\usepackage{graphicx}
\usepackage{bm}
\usepackage{color}
\usepackage{ulem}
\usepackage{physics}
\usepackage{subfigure}
\usepackage{verbatim}
\usepackage{float}
\usepackage{multirow}
\usepackage{diagbox}
\usepackage{soul}
\usepackage[colorlinks,linkcolor=blue,citecolor=blue]{hyperref}

\begin{document}

\title{Tight finite-key analysis for mode-pairing quantum key distribution}

\author{Ze-Hao Wang}
\author{Zhen-Qiang Yin}\email{yinzq@ustc.edu.cn}
\author{Shuang Wang} \email{wshuang@ustc.edu.cn}
\affiliation{CAS Key Laboratory of Quantum Information, University of Science and Technology of China, Hefei, Anhui 230026, China}
\affiliation{CAS Center for Excellence in Quantum Information and Quantum Physics, University of Science and Technology of China, Hefei, Anhui 230026, China}
\affiliation{Hefei National Laboratory, University of Science and Technology of China, Hefei 230088, China}
\author{Rong Wang}
\affiliation{Department of Physics, University of Hong Kong, Pokfulam Road, Hong Kong SAR, China}
\author{Feng-Yu Lu}
\author{Wei Chen}
\author{De-Yong He}
\author{Guang-Can Guo}
\author{Zheng-Fu Han}
\affiliation{CAS Key Laboratory of Quantum Information, University of Science and Technology of China, Hefei, Anhui 230026, China}
\affiliation{CAS Center for Excellence in Quantum Information and Quantum Physics, University of Science and Technology of China, Hefei, Anhui 230026, China}
\affiliation{Hefei National Laboratory, University of Science and Technology of China, Hefei 230088, China}

\date{\today}

\begin{abstract}
\noindent Mode-pairing quantum key distribution (MP-QKD) is a potential protocol that is not only immune to all possible detector side channel attacks, but also breaks the repeaterless rate-transmittance bound without needing global phase locking. Here we analyze the finite-key effect for the MP-QKD protocol with rigorous security proof against general attacks. Moreover, we propose a six-state MP-QKD protocol and analyze its finite-key effect. The results show that the original protocol can break the repeaterless rate-transmittance bound with a typical finite number of pulses in practice. And our six-state protocol can improve the secret key rate significantly in long distance cases.
\end{abstract}

\maketitle

\noindent{\bf INTRODUCTION}

\noindent Quantum key distribution (QKD) \cite{BB84,ekert1991quantum}, whose security is guaranteed by the physical law of quantum mechanics, can share private keys between two authorized partners, Alice and Bob. Such keys can encrypt further communications between Alice and Bob by combining the one-time pad, which has been proven to be information-theoretically secure \cite{shannon1949communication,Molotkov2006CONFERENCESAS}.

However, in practical implementations, the imperfections of practical equipments will create some security loopholes. An eavesdropper, Eve, can steal key information without introducing any signal disturbance by taking advantage of these loopholes. A close examination of hacking strategies indicates that most loopholes exist in the detection part of QKD systems \cite{Makarov2006EffectsOD,zhao2008quantum,fung2007phase,Xu2010ExperimentalDO,Lydersen2010HackingCQ,Gerhardt2011FullfieldIO,qian2018hacking,wei2019implementation}. Considering this situation, measurement-device-independent QKD (MDI-QKD) \cite{lo2012measurement} (see also \cite{braunstein2012side}) protocol is proposed, which can remove all loopholes on the vulnerable detection side. Various theory improvements \cite{ma2012alternative,yu2013three,xu2014protocol,curty2014finite,yin2014mismatched,yin2014reference,yu2015statistical,wang2014simulating,zhou2016making,lu2020efficient,hu2021practical,jiang2021higher,Lu2022Unbalanced} and experiments \cite{liu2013experimental,wang2015phase,yin2016measurement,comandar2016quantum,wang2017measurement,Fan-Yuan2022Robustandadaptable} are achieved in recent years. 

Besides the security, high performance and long transmission distance are the eternal pursuits in the research of QKD. However, because of the inevitable transmission loss in the channel, the MDI-QKD and the other QKD protocols which are point-to-point schemes are upper bounded by the secret key capacity of repeaterless QKD \cite{PhysRevLett.102.050503,takeoka2014fundamental,pirandola2017fundamental}. The PLOB (Pirandola, Laurenza, Ottaviani, and Banchi) bound \cite{pirandola2017fundamental}, for example, restricts the secret key rate $R\leq-\log_2(1-\eta)$, where $\eta$ is the total channel transmittance between Alice and Bob. This bound is approximately a linear function of $\eta$, $R \sim O(\eta)$. For breaking this restriction, an interesting work named twin-field QKD (TF-QKD) was proposed \cite{lucamarini2018overcoming}, whose secret key rate $R \sim O(\sqrt{\eta})$. Moreover, it is an MDI-like protocol that is immune to all detection attacks. Subsequently, many variants of TF-QKD are proposed \cite{ma2018phase,wang2018twin,curty2019simple,cui2019twin,wang2020optimized}, such as sending-or-not-sending QKD (SNS-QKD) \cite{wang2018twin}, phase-matching QKD (PM-QKD) \cite{ma2018phase} and no phase post-selection QKD (NPP-QKD) \cite{cui2019twin}. Because of their high performance, they have attracted much attention and remarkable progress has been made not just in the theory \cite{maeda2019repeaterless,jiang2019unconditional,lu2019practical,xu2020sending,zeng2020symmetry,curras2021tight}, but also in the experimental implementations \cite{minder2019experimental,zhong2019proof,wang2019beating,liu2019experimental,fang2020implementation,chen2020sending,liu2021field,chen2021twin,clivati2022coherent,pittaluga2021600,chen2022quantum,wang2022twin}. However, because TF-type QKD depends on stable interference between coherent states, phase-tracking is needed for compensating the phase fluctuation on the channels while phase-locking is needed for locking the frequency and phase of Alice and Bob's lasers. These challenging techniques significantly increase the complexity of experimental systems and may bring extra noise.

Recently, two works named asynchronous-MDI-QKD \cite{xie2022breaking} and mode-pairing QKD (MP-QKD) \cite{zeng2022quantum} are proposed almost simultaneously. Surprisingly, while their quantum state preparation and measurement are almost the same as the time-bin-phase coding MDI-QKD \cite{ma2012alternative}, the key rate can be increased dramatically just by defining the key bit differently in post processing step. As a result, beating the PLOB bound becomes possible even without the help of phase-locking and phase-tracking. 

For ease of understanding, let's review the MP-QKD in a simple way. Similar with the time-bin-phase coding MDI-QKD \cite{ma2012alternative}, in each round, Alice and Bob send weak coherent pulses with different intensities and phases to Charlie. Then, after Charlie announces each round leads to a single click or not by his interference measurement, Alice and Bob may pair two clicked pulses to a key generation event provided the timing interval of the two paired pulses is smaller than the coherent time of the lasers.
Note that this pairing step is missing in the time-bin-phase coding MDI-QKD \cite{ma2012alternative}, where only two adjacent and clicked optical pulses, i.e. coincidence detection, can be used to generate key bit, thus its key rate must be proportional to the channel transmittance $\eta$. Indeed, this subtle pairing strategy leads to a dramatic increment of the key rate. 

Although an elegant security proof for MP-QKD has been given in Ref. \cite{zeng2022quantum}, its security and performance in non-asymptotic situations are still unknown. Here, we show a finite-key analysis of the MP-QKD protocol. Our analysis applies to coherent attack and satisfies the definition of composable security, namely the secret keys are perfect keys except a failure probability no larger than $\varepsilon_{\rm{\rm{tol}}}$. It's confirmed that PLOB bound can be surpassed in MP-QKD with a moderate number of optical pulses, typically $10^{13}$. Moreover, we propose a six-state MP-QKD protocol that can improve the secret key rate, and show its performance in the finite-key regions. Our results can be directly applied in future practical experiments of MP-QKD.

\hfill

\noindent{\bf RESULTS}

\noindent{\bf Original protocol }

\begin{figure}[ht]
\centering
\includegraphics[scale=0.36]{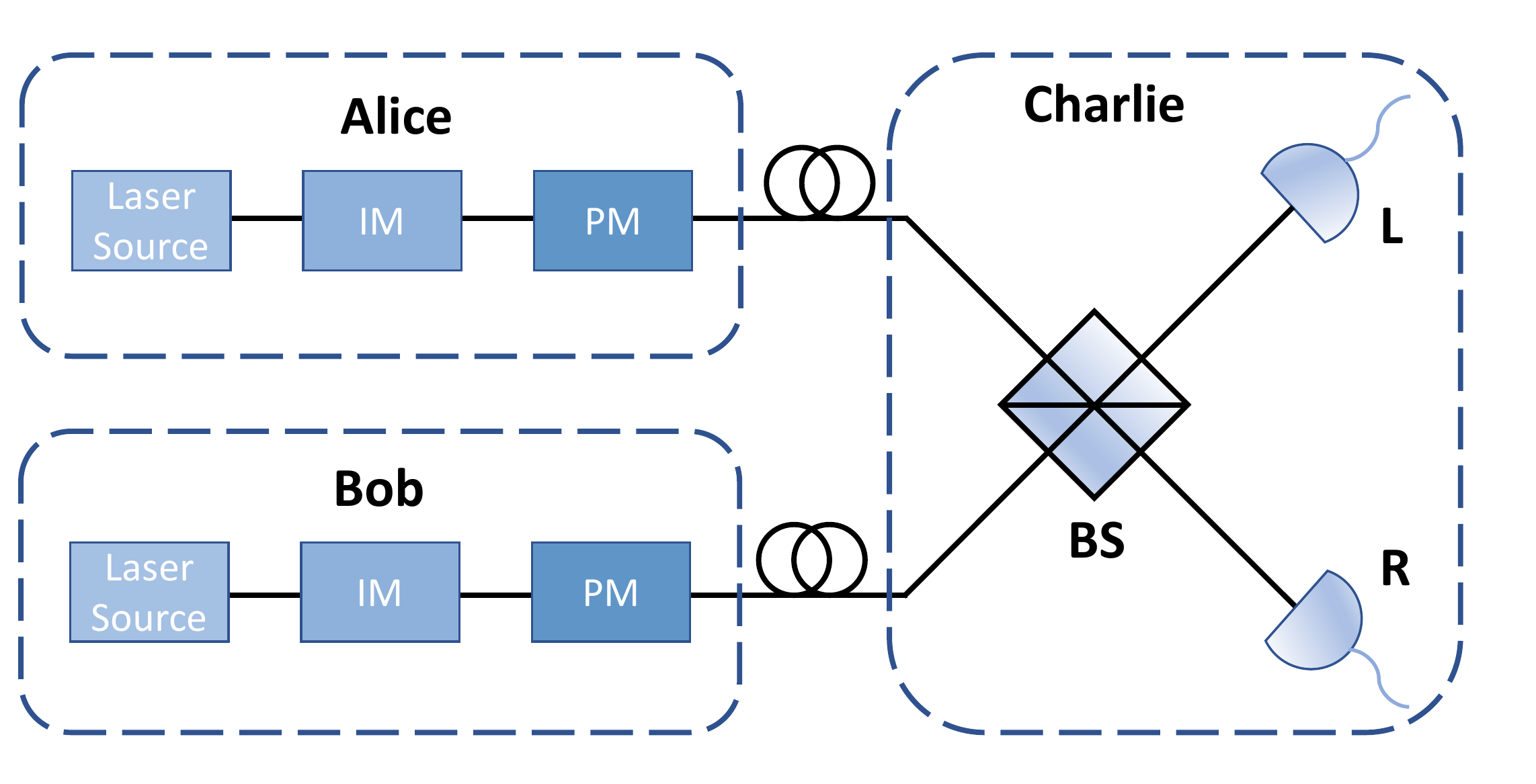}
\caption{Schematic diagram of the MP-QKD protocol. Alice (Bob) prepares weak coherent pulses with random intensities chosen from $\{\mu_a,\nu_a,o\}$ ($\{\mu_b,\nu_b,o\}$) and random phases $\theta^i_a\in[0,2\pi)$ ($\theta^i_b\in[0,2\pi)$), then they send them to an untrusted measurement site, Charlie. According to the effective detection events announced by Charlie, Alice and Bob take the postprocessing to obtain the secret keys. IM, intensity modulation; PM, phase modulation; BS, beam splitter.}
\label{Fig_Schematic_diagram}
\end{figure}

\noindent Here we consider a three-intensity decoy-state scheme \cite{zeng2022quantum}. In this scheme, Alice randomly sends the phase-randomized coherent pulses to Charlie. The intensities of the pulses are chosen from $\{\mu_a,\nu_a,o\}$ $(\mu_a > \nu_a > o=0)$ with probabilities $p_{\mu_a}$, $p_{\nu_a}$ and $p_{o}$, respectively. Bob takes a similar operation to distribute the pulses. According to the effective detection events announced by Charlie, Alice and Bob take the postprocessing to obtain the secret keys. The schematic diagram is illustrated in Fig. \ref{Fig_Schematic_diagram}. And a detailed description of the scheme is presented as follows.

1.\textbf{State preparation}. The first two steps are repeated by Alice and Bob for $N$ rounds to obtain sufficient data. In the $i$-th round ($i\in$ $\{1,2,...,N\}$), Alice prepares a coherent state $|\sqrt{k^i_a} \exp({\rm{i}}\theta^i_a)\rangle$ with probability $p_{k_a}$, where the intensity $k^i_a$ is chosen randomly from the set $\{\mu_a,\nu_a,o\}$ ($\mu_a>\nu_a>o=0$) and the phase $\theta^i_a$ is chosen uniformly from $[0,2\pi)$. Similarly, Bob chooses $k^i_b$ and $\theta^i_b$ then prepares a weak coherent state $|\sqrt{k^i_b} \exp({\rm{i}}\theta^i_b)\rangle$. 

2.\textbf{Measurement}. Alice and Bob send two pulses $|\sqrt{k^i_a} \exp({\rm{i}}\theta^i_a)\rangle$ and $|\sqrt{k^i_b} \exp({\rm{i}}\theta^i_b)\rangle$ to Charlie for interference measurement. After the measurement, Charlie announces the detection results of detectors L and R. If only detector L or R clicks, $C^i=1$. And $C^i=0$ for the other cases. 

\begin{figure}[ht]
\centering
\includegraphics[scale=0.5]{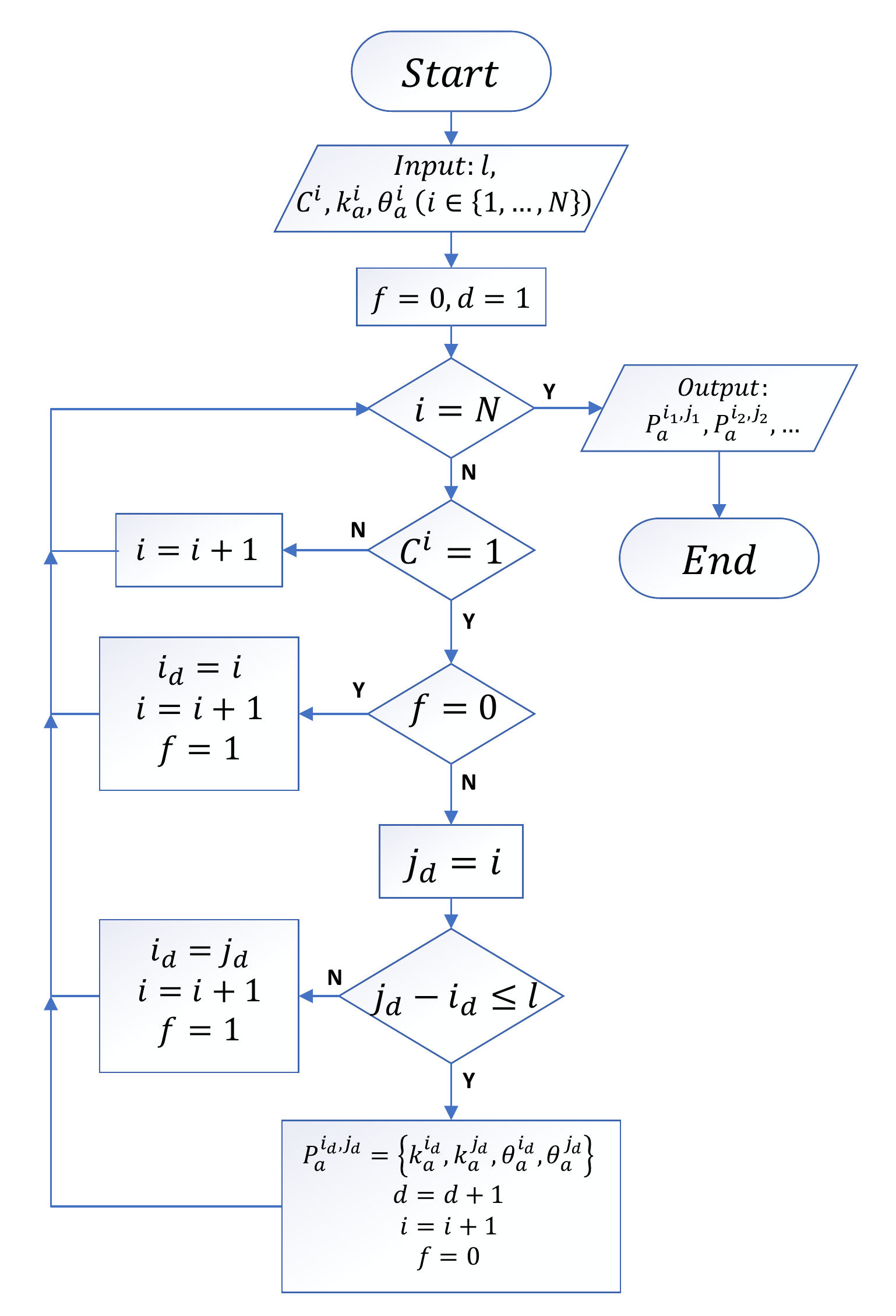}
\caption{Flow chart of Alice's pairing strategy.  At the beginning, Alice inputs the maximal pairing interval $l$, $C^i$, $k_a^i$, and $\theta_a^i$. By employing the pairing strategy, she outputs the data sets $P_a^{i_1,j_1},P_a^{i_2,j_2},...$ }
\label{Fig_Flow chart of the pairing strategy}
\end{figure}

3.\textbf{Mode pairing}. For all rounds with effective detection ($C^i=1$), Alice and Bob employ a strategy to group two effective detection events as a pair. The specific pairing strategy is shown as follows:

Considering that when the time interval between adjacent pulses becomes too large, the key information suffers from phase fluctuation. Alice and Bob define a maximal pairing interval $l$ before the pairing, which denotes the maximal interval pulse number in an effective event pair. $l$ can be estimated by multiplying the laser coherence time by the system repetition rate in practically. Alice (Bob) calculates the interval between the first and second effective detection event. If the interval is less than or equal to $l$, she (he) records an effective event pair $P_{a(b)}^{i_1,j_1}=\{k^{i_1}_{a(b)},k^{j_1}_{a(b)},\theta^{i_1}_{a(b)},\theta^{j_1}_{a(b)}\}$, where $i_1$ and $j_1$ are the round numbers of the first and the second effective detection event, then considers the interval between the third and the fourth effective detection event. Otherwise, the first effective detection event is dropped, then she (he) considers the interval between the second and the third. Until the last effective detection event, she (he) completes the mode pairing. A flow chart of Alice's pairing strategy is presented in Fig. \ref{Fig_Flow chart of the pairing strategy}.

4.\textbf{Basis sifting}. Based on the intensities of two grouped rounds, Alice (Bob) labels the 'basis' of each $P_{a(b)}^{i,j}$ as:

(a) $Z$-basis: if one of the intensities $k^{i}_{a(b)}$ and $k^{j}_{a(b)}$ is 0 and the other is nonzero;

(b) $X$-basis: if $k^{i}_{a(b)}=k^{j}_{a(b)}\neq 0$;

(c) '0'-basis: if $k^{i}_{a(b)}=k^{j}_{a(b)}= 0$;

(d) 'discard': if $0 \neq k^{i}_{a(b)} \neq  k^{j}_{a(b)} \neq 0$.

\begin{table}[ht]
\caption{Alice and Bob's bases assignment.}
\renewcommand{\arraystretch}{1.2}
\begin{tabular}{|l|ccc|}
\hline
\diagbox{$k^j_{a(b)}$}{$k^i_{a(b)}$} & $\mu_{a(b)}$ & $\nu_{a(b)}$ & $o$  \\
\colrule
$\mu_{a(b)}$ & $X$-basis & 'discard' & $Z$-basis \\
$\nu_{a(b)}$ & 'discard' & $X$-basis & $Z$-basis \\
$o$          & $Z$-basis & $Z$-basis & '0'-basis \\
\hline
\end{tabular}
\label{Tab_bases assignment}
\end{table}

See also the Tab. \ref{Tab_bases assignment}.

Then they announce the bases ($Z$-basis, $X$-basis, '0'-basis, or 'discard'), the sum of the intensities $k^i_a+k^j_a$ and $k^i_b+k^j_b$ for each pair of $P_{a}^{i,j}$ and $P_{b}^{i,j}$ respectively. Let's denote the pair of $P_{a}^{i,j}$ and $P_{b}^{i,j}$ by $P^{i,j}$ for simplicity. Then for each $P^{i,j}$, if the announced bases for $P_{a}^{i,j}$ and $P_{b}^{i,j}$ are both $Z$-basis, $X$-basis or '0'-basis, they record $P^{i,j}$ as $Z$-pair, $X$-pair, or '0'-pair respectively; if one of them is '0'-basis and the other one is $X$-basis ($Z$-basis), they record $P^{i,j}$ as $X$-pair ($Z$-pair); if both of the announced bases are '0', they record it as '0'-pair; and otherwise, they discard $P^{i,j}$. See also the Tab. \ref{Tab_pair assignment}.
\begin{table}[ht]
\caption{The pair assignment.}
\renewcommand{\arraystretch}{1.2}
\begin{tabular}{|l|ccc|}
\hline
\diagbox{Bob}{Alice} & $Z$-basis & $X$-basis & '0'-basis \\
\colrule
$Z$-basis & $Z$-pair & 'discard' & $Z$-pair \\
$X$-basis & 'discard' & $X$-pair & $X$-pair \\
'0'-basis & $Z$-pair & $X$-pair & '0'-pair \\
\hline
\end{tabular}
\label{Tab_pair assignment}
\end{table}

5.\textbf{Key mapping}. For each $Z$-pair $P^{i,j}$, if $k^j_a \neq k^i_a=0$, Alice sets her key to $\kappa_a=0$; if $k^i_a \neq k^j_a=0$, Alice sets her key to $\kappa_a=1$; and for the condition $k^i_a=k^j_a=0$, she sets her key $\kappa_a$ to either 0 or 1 at random with a $1/2$ probability. The setting of Bob's key is contrary to Alice's. If $k^i_b \neq k^j_b=0$, Bob sets his key to $\kappa_b=0$; if $k^j_b \neq k^i_b=0$, Bob sets his key to $\kappa_b=1$; and for the condition $k^i_b=k^j_b=0$, he sets his key $\kappa_b$ to either 0 or 1 at random with a $1/2$ probability.


For each $X$-pair $P^{i,j}$, Alice's key is extracted from the relative phase $\theta_a:=(\theta^j_a-\theta^i_a)  \mod  2\pi$, where the raw key bit is $\kappa_a=\lfloor \theta_a/\pi \rfloor$ and the alignment angle is $\delta_a=\theta_a  \mod  \pi$ $\left(\delta_a\in[0,\pi)\right)$. Bob also calculates his raw key bit $\kappa_b$ and alignment angle $\delta_b$ in the same way. Then they announce the alignment angles $\delta_a$ and $\delta_b$. If $|\delta_a-\delta_b| \leq \Delta$, they keep $P^{i,j}$; if $|\delta_a-\delta_b| \geq \pi-\Delta$, Bob flips the raw key bit $\kappa_b$ and they keep $P^{i,j}$; otherwise, they discard them. Moreover, if Charlie's clicks are (L,R) or (R,L), Bob flips $\kappa^b$. It should be noted that if $P^{i,j}$ consists of both '0'-basis and $X$-basis, they retain all the data pairs whatever the value of $\delta_a$ and $\delta_b$ are. We define three sets $\mathcal{Z}$, $\mathcal{X}$ and $\mathcal{V}$ here, which include all $Z$-pair, $X$-pair, and '0'-pair $P^{i,j}$ respectively.

6.\textbf{Parameter estimation}. Alice and Bob choose $P^{i,j}$ satisfying $(k^i_a+k^j_a,k^i_b+k^j_b)\in\{(\mu_a,\mu_b),(\mu_a,\nu_b),(\nu_a,\mu_b),(\nu_a,\nu_b)\}$ in the set $\mathcal{Z}$ to form the $n_Z$-length raw key bit strings $\mathbf{Z}$ and $\mathbf{Z'}$, respectively. And through the decoy-state method, Alice and Bob estimate the lower bound of the single-photon effective detection number in the raw key, $n_{Z_1}^{\rm{L}}$, according to the events in $\mathcal{Z}$ and $\mathcal{V}$. The upper bound of the single-photon phase error rate of $n_{Z_1}^{\rm{L}}$, $e_{Z_1}^{\rm{ph,U}}$, is estimated by the events in $\mathcal{X}$ and $\mathcal{V}$.

7.\textbf{Error correction}. Alice and Bob exploit an information reconciliation scheme to correct $\mathbf{Z'}$, in which Bob acquires an estimation $\hat{\mathbf{Z}}$ of $\mathbf{Z}$ from Alice. This process reveals at most $\lambda_{\rm{EC}}$ bits Alice's raw key. Then, for verifying the success of error correction, Alice employs a random universal hash function to compute a hash of $\mathbf{Z}$ of length $\log_2(2/\varepsilon_{\rm{cor}})$, and sends the hash and hash function to Bob. If the hash computed by Bob is the same as Alice's, $\Pr\left(\mathbf{Z}\neq\hat{\mathbf{Z}}\right)\leq\varepsilon_{\rm{cor}}$, they proceed this protocol. Otherwise, the protocol aborts.

8.\textbf{Private amplification}. Alice and Bob exploit a privacy amplification scheme based on two-universal hashing \cite{renner2008security} to extract two $l_{o}$-length bit strings $\mathbf{S}$ and $\hat{\mathbf{S}}$ from $\mathbf{Z}$ and $\hat{\mathbf{Z}}$ respectively. $\mathbf{S}$ and $\hat{\mathbf{S}}$ are the secret keys.

Then, we show one of the main results of our paper. If the error correction step is passed, the protocol is $\varepsilon_{\rm{cor}}$-correct. And the protocol is $\varepsilon_{\rm{sec}}$-secret if the secret key length
\begin{equation}
\begin{aligned}
l_o \leq& n_{Z_1}^{\rm{L}}\left[ 1-h(e_{Z_1}^{\rm{ph,U}})\right]-{\lambda_{\rm{EC}}}\\
&-\log_2\frac{2}{\varepsilon_{\rm{cor}}}-2\log_2 \frac{1}{\sqrt{2}\hat{\varepsilon}\varepsilon_{\rm{PA}}}, 
\end{aligned}
\end{equation}
where $h(x)=-x\log_2 x -(1-x) \log_2 (1-x)$ is the binary Shannon entropy function, and $\varepsilon_{\rm{cor}}$, $\hat{\varepsilon}$, and $\varepsilon_{\rm{PA}}$ are the security coefficients which the users may optimize over. $n_{Z_1}^{\rm{L}}$ is the lower bound of the single-photon effective detection number in the raw key with a failure probability $\varepsilon_{1}$. $e_{Z_1}^{\rm{ph,U}}$ is the upper bound of the single-photon phase error rate of $n_{Z_1}^{\rm{L}}$ with a failure probability $\varepsilon_e$. $n_{Z_1}^{\rm{L}}$ and $e_{Z_1}^{\rm{ph,U}}$ are estimated by the decoy-state method, which is shown in the Supplementary Note C. And ${\lambda_{\rm{EC}}}=f n_Z h (E_Z)$ is the information revealed in the error correction step, where $f$ is the error correction efficiency which is related to the specific error correction scheme, $n_{Z}$ is the length of the raw key and $E_Z$ is the bit-flip error rate between strings $\mathbf{Z}$ and $\mathbf{Z}'$.


\hfill\\

\noindent{\bf Six-state MP-QKD protocol}

\noindent In the original MP-QKD protocol, we bound the Eve's smooth min-entropy $H_{\rm{min}}^{\overline{\varepsilon}} (\mathbf{Z}_1|\mathbf{Z}_{\rm{zm}}E)$ by the error rate of key bits when hypothetical $\sigma_X$ measurements are performed, where $E$ is the auxiliary of Eve before error correction, and $\mathbf{Z}_1,\mathbf{Z}_{\rm{zm}}$ are the corresponding bit strings due to the single-photon and the other events respectively. It's well known that if one can additionally obtain the error rate under $\sigma_Y$ measurements, the min-entropy will be estimated more tightly and higher key rate is expected, e.g. six-state protocol outperforms the original BB84 in most cases.

Surprisingly, it's possible to obtain both the error rates under hypothetical $\sigma_X$  and $\sigma_Y$ measurements in the MP-QKD. Let's explain this point intuitively. When the pair $P^{i,j}_a$ happens to be a $Z$-basis and also a single-photon event, its hypothetical $\sigma_X$  and $\sigma_Y$ measurements just lead to quantum superpositions $(|i\rangle \pm |j\rangle)/\sqrt{2}$ or $(|i\rangle \pm {\rm{i}}|j\rangle)/\sqrt{2}$ respectively. The quantum state $|i\rangle$($|j\rangle$) denote that in the i-th and j-th rounds, only single-photon is in the $i-th$ ($j-th$) round. Note that if $P^{i,j}_a$ happens to be an $X$-basis and also a single-photon event, $(|i\rangle \pm e^{{\rm{i}}\theta_a} |j\rangle)/\sqrt{2}$ will be prepared, in which  $\theta_a$ ranges from $[0,\pi)$. Equivalently,  $(|i\rangle \pm e^{{\rm{i}}\theta_a} |j\rangle)/\sqrt{2}$ can be rewritten as  $(|i\rangle \pm e^{{\rm{i}}\theta_a} |j\rangle)/\sqrt{2}$ if $\theta_a\in[0,\pi/2)$, and $(|i\rangle \pm {\rm{i}} e^{{\rm{i}}\theta'_a} |j\rangle)/\sqrt{2}$ if $\theta'_a\in[\pi/2,\pi)$, where $\theta'_a:=\theta_a \mod (\pi/2)$.  Just as mentioned in Supplemental Note, $\theta'_a$ can be resulted by Alice's unitary operation on her local qubits, thus has no effect on the security. This implies that for any $\theta'_a$,  the former one and latter may be used to estimate the error rates under  $\sigma_X$  and $\sigma_Y$ measurements respectively,  so $X$-basis is indeed $XY$-basis in this sense. Finally, a six-state like MP-QKD is possible.
Following this idea, we present a detailed description of this scheme and the results of security proof as follows:

1.\textbf{State preparation}. Same as the step 1 in the original protocol. 

2.\textbf{Measurement}. Same as the step 2 in the original protocol.

3.\textbf{Mode pairing}. Same as the step 3 in the original protocol.

4.\textbf{Basis sifting}. Based on the intensities of two grouped rounds, Alice (Bob) labels the 'basis' of $P_{a(b)}^{i,j}$ as:

(a) $Z$-basis: if one of the intensities $k^{i}_{a(b)}$ and $k^{j}_{a(b)}$ is 0 and the other is nonzero;

(b) $XY$-basis: if $k^{i}_{a(b)}=k^{j}_{a(b)}\neq 0$;

(c) '0'-basis: if $k^{i}_{a(b)}=k^{j}_{a(b)}= 0$;

(d) 'discard': if $0 \neq k^{i}_{a(b)} \neq  k^{j}_{a(b)} \neq 0$.

Then they announce the basis ($Z$-basis, $XY$-basis, '0'-basis or 'discard') and the sum of the intensities $k^i_a+k^j_a$ and $k^i_b+k^j_b$ for each pair of $P_{a}^{i,j}$ and $P_{b}^{i,j}$ respectively. Let's denote the pair of $P_{a}^{i,j}$ and $P_{b}^{i,j}$ by $P^{i,j}$ for simplicity. Then for each $P^{i,j}$: if the announced bases for $P_{a}^{i,j}$ and $P_{b}^{i,j}$ are both $Z$-basis, $XY$-basis or '0'-basis, they record $P^{i,j}$ as $Z$-pair, $XY$-pair, or '0'-pair respectively; if one of them is '0'-basis and the other one is $XY$-basis ($Z$-basis), they record $P^{i,j}$ as $XY$-pair ($Z$-pair); if both of the announced bases are '0', they record it as '0'-pair; and otherwise, they discard $P^{i,j}$.

5.\textbf{Key mapping}. The key mapping step of $Z$-pair $P^{i,j}$ is the same as the step in the original protocol. For each $Z$-pair $P^{i,j}$, if $k^j_a \neq k^i_a=0$, Alice sets her key to $\kappa_a=0$; if $k^i_a \neq k^j_a=0$, Alice sets her key to $\kappa_a=1$; and for the condition $k^i_a=k^j_a=0$, she sets her key $\kappa_a$ to either 0 or 1 at random with a $1/2$ probability. The setting of Bob's key is contrary to Alice's. If $k^i_b \neq k^j_b=0$, Bob sets his key to $\kappa_b=0$; if $k^j_b \neq k^i_b=0$, Bob sets his key to $\kappa_b=1$; and for the condition $k^i_b=k^j_b=0$, he sets his key $\kappa_b$ to either 0 or 1 at random with a $1/2$ probability.

\begin{table}[ht] 
\caption{The rules of the flip operation to Bob's basis $r_b$ and bit $\kappa_b$. The word "flip" means Bob flips the value, and the word "no" means Bob just keeps the value.}
\renewcommand{\arraystretch}{1.3}
\begin{ruledtabular} 
\begin{tabular}{l|cc|cc}
 & \multicolumn{2}{l|}{$r_a=0,r_b=0$} & \multicolumn{2}{l}{$r_a=1,r_b=1$}  \\
 & $r_b$ & $\kappa_b$ &$r_b$ & $\kappa_b$ \\
\colrule
$|\delta_a-\delta_b|\leq\Delta$   & no & no & no & no  \\
\hline
\hline
 & \multicolumn{2}{l|}{$r_a=0,r_b=1$} & \multicolumn{2}{l}{$r_a=1,r_b=0$}  \\
 & $r_b$ & $\kappa_b$ &$r_b$ & $\kappa_b$ \\
\colrule
$\delta_b-\delta_a \geq \pi/2-\Delta$   & flip & flip & flip & no  \\
$\delta_a-\delta_b \geq \pi/2-\Delta$   & flip & no & flip & flip 
\end{tabular}
\end{ruledtabular}
\label{Tab_rules of the reverse operation}
\end{table}

For each $XY$-pair $P^{i,j}$, Alice's key and basis are extracted from the relative phase $\theta_a=(\theta^j_a-\theta^i_a) \mod 2\pi$, where the raw key bit is $\kappa_a=\lfloor \theta_a/\pi \rfloor$, the alignment angle is $\delta_a=\theta_a  \mod  (\pi/2)$ $\left(\delta_a\in[0,\pi/2)\right)$, and $r_a=\lfloor \theta_a/(\pi/2) \rfloor -2 \kappa_a$, where $r_a=0$ and $1$ denotes $X$- and $Y$-bases, respectively. Bob also calculates his raw key bit $\kappa_b$, alignment angle $\delta_b$, and obtains the information of the basis $r_b$ in the same way. Then they announce the alignment angles $\delta_a$, $\delta_b$ and the values $r_a$ and $r_b$.
If $|\delta_a-\delta_b| \leq \Delta$ and $r_a=r_b$, they keep $P^{i,j}$ and label them as $X$-pair if $r_a=0$ or $Y$-pair if $r_a=1$; 
if $|\delta_a-\delta_b| \geq \pi/2-\Delta$ and $r_a \neq r_b$, Bob may flip the basis $r_b$ and the raw key bit $\kappa_b$ according to the rules in the Tab. \ref{Tab_rules of the reverse operation}, then keep $P^{i,j}$ and label them as $X$-pair if $r_a (r_b)=0$ or $Y$-pair if $r_a(r_b)=1$; otherwise, they discard them. Moreover, if Charlie's clicks are (L,R) or (R,L), Bob flips $\kappa^b$. We define three sets $\mathcal{Z}$, $\mathcal{XY}$ and $\mathcal{V}$ here, which include all reserved $Z$-pair, $XY$-pair, and '0'-pair $P^{i,j}$ respectively.

6.\textbf{Parameter estimation}. Alice and Bob choose $P^{i,j}$ satisfying $(k^i_a+k^j_a,k^i_b+k^j_b)\in\{(\mu_a,\mu_b),(\mu_a,\nu_b),(\nu_a,\mu_b),(\nu_a,\nu_b)\}$ in the set $\mathcal{Z}$ to form the $n_Z$-length raw key bit strings $\mathbf{Z}$ and $\mathbf{Z'}$, respectively. And through the decoy-state method, Alice and Bob estimate the number of effective single-photon events $n_{Z_1}^{\rm{L}}$ according the events in $\mathcal{Z}$ and $\mathcal{V}$.

7.\textbf{Error correction}. Same as the step 7 in the original protocol.

8.\textbf{Private amplification}. Same as the step 8 in the original protocol.

Similar with the analysis of the original protocol, we can bound the secret key length with
\begin{equation}
\begin{aligned}
l_{s} \leq& n_{Z_1}^{\rm{L}}\left(1-e_{Z_1}^{\rm{bit,U}}\right)\\
&\times \left[1-h\left(\frac{1-\frac{1}{2}\left(e_{Z_1}^{\rm{bit,U}}+\left(e_{X_1}^{\rm{bit}}+e_{Y_1}^{\rm{bit}}\right)^{\rm{U}}\right)}{1-e_{Z_1}^{\rm{bit,U}}}\right)\right]\\
&-{\lambda_{\rm{EC}}}-\log_2 \frac{2}{\varepsilon_{\rm{cor}}}-2\log_2{\frac{1}{\sqrt{2}\hat{\varepsilon}\varepsilon_{\rm{PA}}}},
\end{aligned}
\end{equation}
where $e_{Z_1}^{\rm{bit,U}}$ is the upper bound of the single-photon bit error rate of $n_{Z_1}^{\rm{L}}$ with a failure probability $\varepsilon_e'$. $\left(e_{X_1}^{\rm{bit}}+e_{Y_1}^{\rm{bit}}\right)^{\rm{U}}$ is the upper bound of the sum of the single-photon bit error rate of $n_{Z_1}^{\rm{L}}$ if Alice and Bob take $\sigma_X$ or $\sigma_Y$ measurement, with a failure probability $\varepsilon_e''$.The detailed security proof is shown in Supplementary Note A. And the specific calculation is shown in Supplementary Note C.

\hfill\\
\noindent{\bf Discussion}

\noindent In this section, we numerically simulate the secret key rate of the original MP-QKD and the six-state MP-QKD protocol in finite-key size cases and analyze the results. Partial experimental parameters which are fixed in the following simulations are listed in Tab. \ref{Tab_experimental parameters}. Some other parameters, including the maximal pairing interval, $l$, and the total pulse numbers, $N$, are given in the captions of figures below. It should be noted that in Refs. \cite{xie2022breaking,zeng2022quantum}, there are all specific analyses about the misalignment-error. The misalignment-error will increase as $l$ increases, and will decrease as the system frequency increases. Here we omit the specific analysis and just take $e_d^Z=0.5\%$, $e_d^X=5\%$. As is shown in Ref. \cite{zhu2022experimental,zhou2022experimental}, it is realizable to take a key distribution with $l=10^4$, $e_d^X=0.5\%$, and $e_d^X=5\%$ by employing a 4 GHz system \cite{wang2022twin}. If we improve the performance of the lasers, $l=10^6$ is achievable. In the following analysis, we focus on $l=10^2$, $l=10^4$, and $l=10^6$. They correspond to three different performance systems.

Meanwhile, we optimize the intensities and their corresponding sending probabilities for each distance in the simulation. Without loss of generality, we assume that the distance between Alice and Charlie, $L_a$, and the distance between Bob and Charlie, $L_b$, are the same. Under this situation, $\mu=\mu_a=\mu_b$, $\nu=\nu_a=\nu_b$, $p_{\mu}=p_{\mu_a}=p_{\mu_b}$, and $p_{\nu}=p_{\nu_a}=p_{\nu_b}$. Then, there are only five parameters, $\mu$, $\nu$, $p_\mu$, $p_\nu$, and $\Delta$, needed to be optimized. 

\begin{table}[ht]
\caption{\label{Tab_experimental parameters} Experimental parameters used in the numerical simulations. Here, $p_d$ is the dark counting rate per pulse of Charlie's detectors; $\eta_d$ is the detection efficiency of Charlie's detectors; $\alpha$ is the fiber loss coefficient (dB/km); $f$ is  the error-correction efficiency; $\varepsilon_{\rm{tol}}$ is the total secure coefficient. $e_d^Z$ and $e_d^X$ are the misalignment-error of the sets $\mathcal{Z}$ and $\mathcal{X}(\mathcal{XY})$, respectively.} 
\renewcommand{\arraystretch}{1.2}
\begin{ruledtabular} 
\begin{tabular}{lcccccc}
$p_d$ & $\eta_d$ & $\alpha$ & $f$ & $\varepsilon_{\rm{tol}}$ & $e_d^Z$ & $e_d^X$ \\
\colrule
$1 \times 10^{-8}$ & $70\%$ & $0.2$ & $1.1$ & $1\times10^{-10}$ & $0.5\%$ & $5\%$
\end{tabular}
\end{ruledtabular}
\end{table}

\begin{figure}[ht]
\includegraphics[scale=0.63]{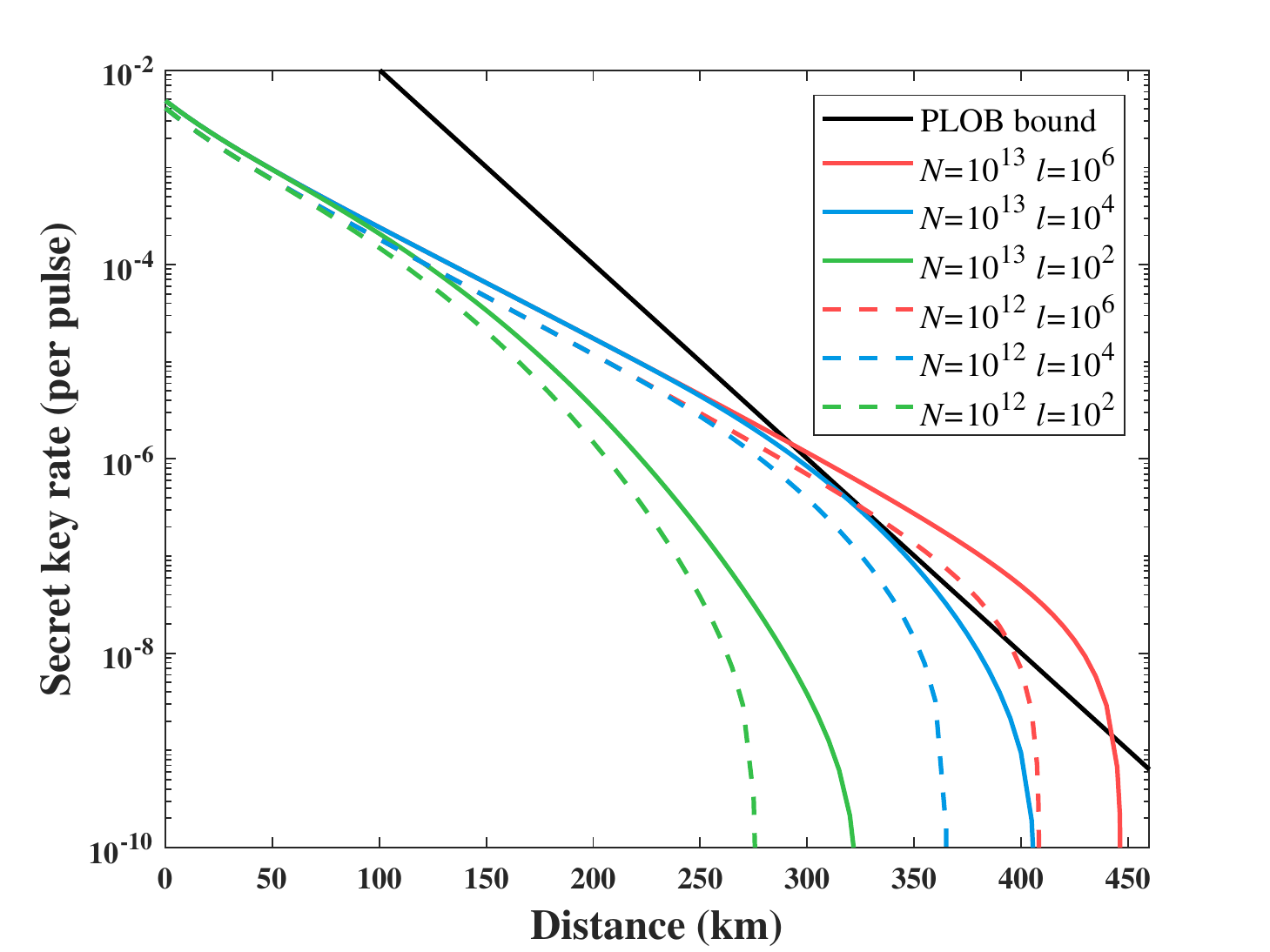}
\caption{Comparison of the secret key rate (per pulse) of the original protocol among two different total pulse numbers, $N$, and three different maximal pairing intervals, $l$, as a function of transmission distance (the distance between Alice and Bob). The red, blue, and green lines represent the secret key rates of the origin MP-QKD protocol under the condition that $l=10^6$, $l=10^4$, and $l=10^2$, respectively. The solid lines and dash lines represent the secret key rates under the condition $N=10^{13}$ and $N=10^{12}$, respectively. The black line is the PLOB bound.}
\label{Fig_simulation_origin}
\end{figure}

\begin{figure*}[ht]
    \includegraphics[scale=0.8]{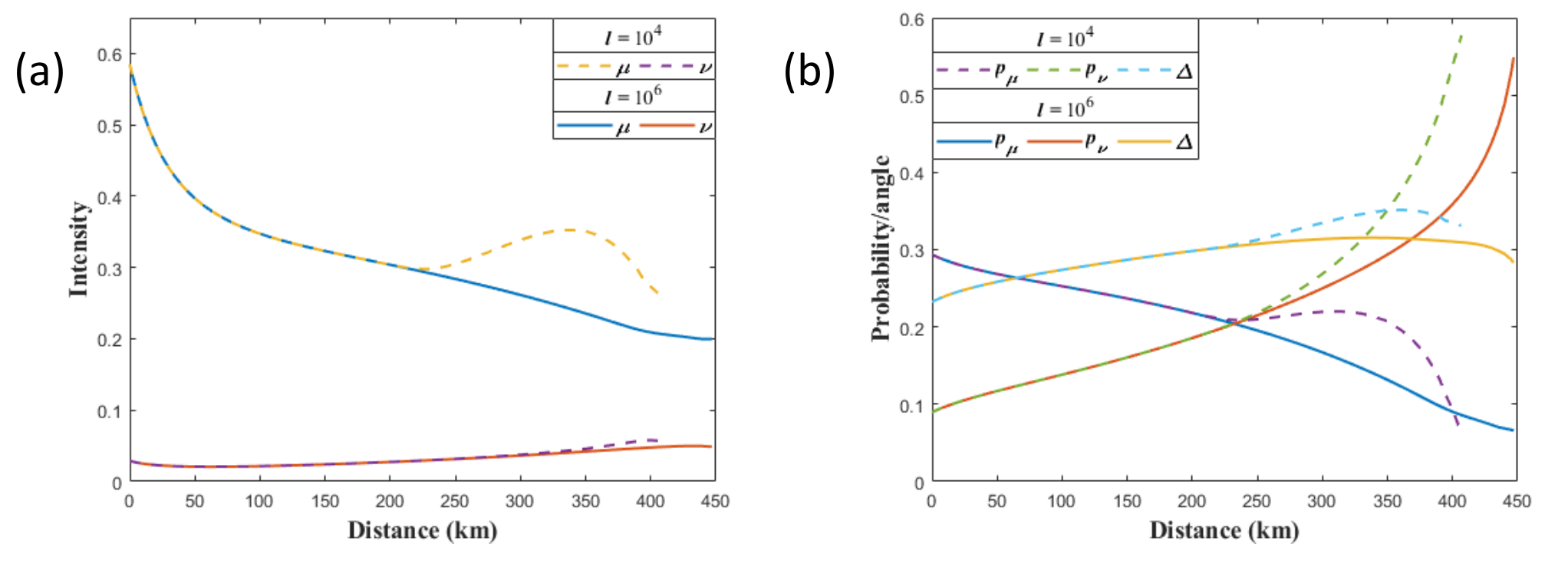}
    \caption{Optimized variables evolution over increasing distance when the number of total pulses is $10^{13}$. The solid lines and dash lines represent the secret key rates under the condition $l=10^{6}$ and $l=10^{4}$, respectively.}
    \label{Fig_optimized_parameter}
\end{figure*}

\begin{figure}[htb]
\includegraphics[scale=0.57]{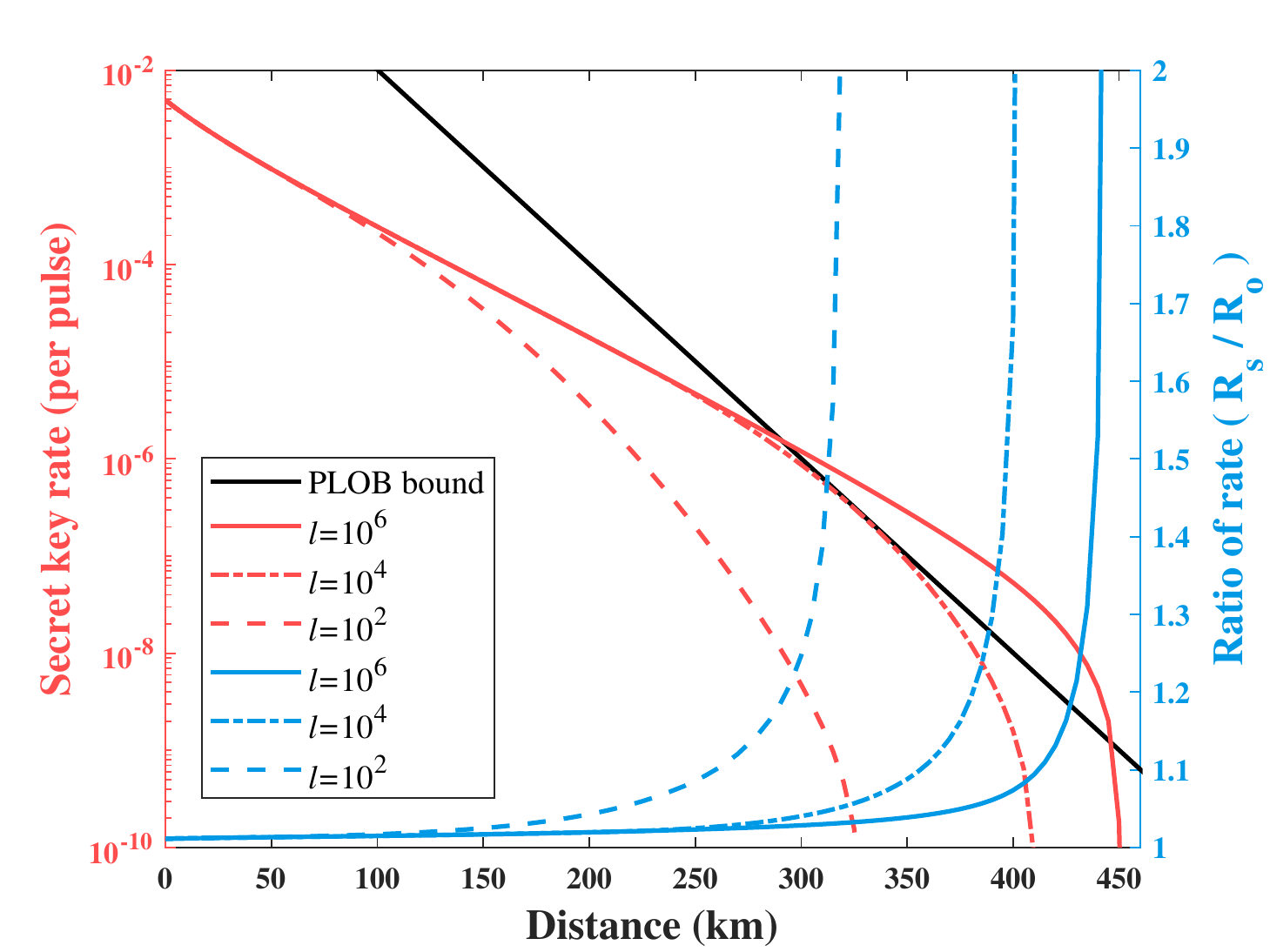}
\caption{The secret key rate (per pulse) of the six-state protocol and the ratio of $R_s$ and $R_o$ as a function of transmission distance (the distance between Alice and Bob). Besides the fixed parameters listed in Tab. \ref{Tab_experimental parameters}, $N=10^{13}$. The ordinate of red lines is the secret key rate (per pulse), and the ordinate of blue lines is the ratio of $R_s$ and $R_o$ ($R_s/R_o$). The solid, dash-dotted, and dash lines represent the secret key rate of the six-state MP-QKD protocol under the condition that $l=10^6$, $l=10^4$, and $l=10^2$, respectively. The black line is the PLOB bound.}
\label{Fig_simulation_six_state}
\end{figure}

With the experimental parameters and optimized parameters, we simulate observed values first, including effective detection number and bit error number. The specific formulas are shown in Supplementary Note B. Then, with those observed values, we can obtain $n_{Z_1}^{\rm{L}}$, $e_{Z_1}^{\rm{ph,U}}$, $e_{Z_1}^{\rm{bit,U}}$, and $(e_{X_1}^{\rm{bit}}+e_{Y_1}^{\rm{bit}})^{\rm{U}}$ by employing the decoy-state method. The specific formulas are shown in Supplementary Note C. Here, we set the same failure probability parameters for simplicity, $\varepsilon_{\rm{cor}}=\hat{\varepsilon}=\varepsilon_{\rm{PA}}=\xi$, $\varepsilon_1=8\xi$, and $\varepsilon_e=\varepsilon_e'=\varepsilon_e''=5\xi$. For a fair comparison, we compare the secret key rate per sending pulse, $R_{o(s)}=l_{o(s)}/N$.

Figure \ref{Fig_simulation_origin} is the simulation of the original MP-QKD protocol with finite-key size analysis. We compare the secret key rate (per pulse) under different total pulse numbers $N$ and different maximal pairing intervals $l$. It indicates that the original protocol can reach up to 446 km under the condition that $N=10^{13}$ and $l = 10^6$. Moreover, when $N=10^{13}$ and $l=10^4$, this protocol is close to the PLOB bound at a distance about 320 km. That means that it can break the PLOB bound if $l > 10^4$ when $N=10^{13}$, or $N > 10^{13}$ when $l=10^{4}$. The secret key rate under a short distance is almost invariable with the increase of the maximal pairing interval. 

For a better understanding of the MP-QKD protocol, we show the evolution of optimized variables over increasing distance when the number of total pulses is ${10^{13}}$ in Fig. \ref{Fig_optimized_parameter}. It can be shown that for $l=10^{6}$, the intensity $\mu$ and its probability $p_\mu$ decrease with the increase of the distance. This is because that with the increase of the distance, the quantum bit error rate (QBER) caused by the multiple-photon term will increase, causing the smaller $\mu$, which is similar to Supplementary Note 5 of Ref. \cite{zeng2022quantum}. And because the statistical fluctuation effect intensifies, the probability $p_\nu$ should be improved to blunt the effect, causing the bigger $p_\nu$ and smaller $p_\mu$. Different from the other protocols, MP-QKD needs enough intensity $o$ to form a $Z$-pair, so $p_o$ is high in our optimization. For $l=10^4$, the parameters are the same as $l=10^6$ when the distance is below 210 km. This is because the gain is high enough under this situation, the quantity of successful pairing can't increase with $l$. Then, with the improvement of the distance, $\mu$ and $p_\mu$ increase and then decrease. These are two tradeoffs between the coincidence count and the QBER, the coincidence count and the statistical fluctuation, respectively. Moreover, $\Delta$ varies between $\pi/16$ and $\pi/8$.

Figure \ref{Fig_simulation_six_state} is the simulation of the six-state MP-QKD protocol with finite-key size analysis. We present the secret key rate under three different maximal pairing intervals. Moreover, we compare the secret key rates of the original protocol and the six-state one, the blue lines show the ratio of $R_s$ to $R_o$. The improvement of our six-state protocol becomes significant in long distance cases.

In conclusion, we prove the composable security of the original MP-QKD protocol in the finite-key regime against general attacks. Methodologically, by employing the uncertainty relation of smooth min- and max-entropy and proving the relation between the single-photon phase error of the $Z$-basis and the single-photon bit error of the $X$-basis, the secret key rate was obtained. Moreover, we propose a six-state MP-QKD protocol and analyze its finite-key effect. The six-state one can obtain a higher secret key rate than the original protocol. This improvement is achieved by a subtle analysis of experimental data, thus any modification of experimental system is not necessary. As shown in the numerically simulation, the MP-QKD protocol can break the PLOB bound when the total pulse numbers is $10^{13}$ and the maximal pairing interval is greater than $10^4$. It can reach up to $446$ km with $10^6$ pairing interval and $10^{13}$ total pulses. Moreover, the simulation shows that the six-state protocol can obtain higher secret key rate than the original protocol, especially on a long distance. Our work provides theoretical tools for key rate calculation in future MP-QKD implementations. 

In this work, our analysis is based on continuous phase randomization , which might be difficult to realize in the real world. For future work, an analysis based on a discrete set is meaningful \cite{cao2015discrete}. And the data of unmatched basis are wasted in our work, the utilization of them may improve the performance or increase the security, similar to some other protocols \cite{laing2010reference,tamaki2014loss,braunstein2012side,PhysRevA.78.042316}. This is also a possible direction for future work.

\hfill

\noindent{\bf METHODS}

\noindent{\bf Secret key rate of the original protocol}

\noindent The sketch of the security proof is presented here while the details are given in Supplementary Note A. We employ the universally composable framework \cite{renner2008security,muller2009composability} to analyze the security of the protocol.  
If the error correction step is passed, the protocol is $\varepsilon_{\rm{cor}}$-correct. And the protocol is $\varepsilon_{\rm{sec}}$-secret if the length of the extracted secret key does not exceed a certain length in the private amplification step. In particular, a protocol is $\varepsilon_{\rm{tol}}$-secure if it is $\varepsilon_{\rm{cor}}$-correct and $\varepsilon_{\rm{sec}}$-secret, $\varepsilon_{\rm{cor}}+\varepsilon_{\rm{sec}} \leq \varepsilon_{\rm{tol}}$.

According to the quantum leftover hash lemma \cite{renner2008security,tomamichel2011leftover}, an $\varepsilon_{\rm{sec}}$-secret key of length $l_o$ can be extracted from the bit string $\mathbf{Z}$ by applying privacy amplification with two-universal hashing, and
\begin{equation}
\begin{aligned}
2\varepsilon +\frac{1}{2}\sqrt{2^{l_o-H_{\rm{min}}^\varepsilon(\mathbf{Z}|E')}} \leq \varepsilon_{\rm{sec}}, 
\end{aligned}
\end{equation}  
where $E'$ is the auxiliary of Eve after error correction, $H_{\rm{min}}^\varepsilon(\mathbf{Z}|E')$ is the conditional smooth min-entropy, which is employed to quantify the average probability that Eve guesses $\mathbf{Z}$ correctly with $E'$ \cite{konig2009Operational}. Because $\lambda_{\rm{EC}}+\log_2 (2/\varepsilon_{\rm{cor}})$ bits are published in the error correction step, the conditional smooth min-entropy $H_{\rm{min}}^\varepsilon(\mathbf{Z}|E')$ can be bounded by
\begin{equation}
\begin{aligned}
H_{\rm{min}}^\varepsilon(\mathbf{Z}|E') \geq H_{\rm{min}}^\varepsilon(\mathbf{Z}|E)- \lambda_{\rm{EC}}-\log_2 \frac{2}{\varepsilon_{\rm{cor}}},
\end{aligned}
\end{equation} 
where $E$ is the auxiliary of Eve before error correction. Moreover,  since the phases of coding states in $Z$-pair $P^{i,j}$  are never revealed, the coding states can be treated as a mixture of Fock states. 
This implies $\mathbf{Z}$ can be decomposed into $\mathbf{Z}_1\mathbf{Z}_{\rm{zm}}$, which are the corresponding bit strings due to the single-photon and the other events. By employing a chain-rule inequality for smooth entropies \cite{vitanov2013chain}, we have 
\begin{equation}
\begin{aligned}
H_{\rm{min}}^\varepsilon(\mathbf{Z}|E) \geq H_{\rm{min}}^{\overline{\varepsilon}} (\mathbf{Z}_1|\mathbf{Z}_{\rm{zm}}E)-2\log_2\frac{\sqrt{2}}{\hat{\varepsilon}},
\end{aligned}
\end{equation}
where $\varepsilon=2\overline{\varepsilon}+\hat{\varepsilon}$. 

The essential of the security proof is how to bound $ H_{\rm{min}}^{\overline{\varepsilon}} (\mathbf{Z}_1|\mathbf{Z}_{\rm{zm}}E)$. We give the sketch of the proof and leave the details in Supplementary Note A. 
By describing the protocol as an entanglement-based one,  Alice (Bob) decides the basis of $P^{i,j}_a$ ($P^{i,j}_b$) and key bit by measuring her (his) local quantum memories. Specifically, $\mathbf{Z}_1$ can be seen as the results of $\sigma_Z$ measurements on Alice's local qubits of single-photon events. Then according to the uncertainty relation for smooth entropies,  $ H_{\rm{min}}^{\overline{\varepsilon}} (\mathbf{Z}_1|\mathbf{Z}_{\rm{zm}}E)$ can be upper-bounded by the phase error rate, i.e. the error rate of key bits generated by the hypothetical $\sigma_X$ measurements on these local qubits.  As an example, when the pair $P^{i,j}_a$ happens to be a $Z$-basis and also a single-photon event,  $\sigma_Z$ measurement on Alice's local qubits makes this single-photon collapse into the position $i$ or $j$, i.e. $|i\rangle$ or $|j\rangle$ respectively. Meanwhile, its hypothetical $\sigma_X$  measurement just leads to quantum super positions $(|i\rangle \pm |j\rangle)/\sqrt{2}$. Actually, one can easily imagine that $(|i\rangle \pm e^{{\rm{i}}\theta_a}|j\rangle)/\sqrt{2}$ will be prepared if $P^{i,j}_a$ is an $X$-basis event. Intuitively, the phase $\theta_a$ may be removed without compromising the security, since this phase may be resulted by  a unitary operation on Alice's local qubits. Then the sketch of the security proof is clear. 

In Supplementary Note A, we prove that the phase error rate for any individual single-photon $Z$-pair $P^{i,j}$ must be equal to the error rate if single-photon $P^{i,j}$ happens to be $X$-pair. Then in non-asymptotic cases, the phase error rate for the raw key string $\mathbf{Z}_1$ has been sampled by the error rate between $\mathbf{X}_1$ and $\mathbf{X}_1'$, where $\mathbf{X}_1$ and $\mathbf{X}_1'$ represent the key strings obtained from single-photon events in $\mathcal{X}$. Finally, we have
\begin{equation}
\begin{aligned}
H_{\rm{min}}^{\overline{\varepsilon}} (\mathbf{Z}_1|\mathbf{Z}_{\rm{zm}}E) \geq& n_{Z_1}^{\rm{L}}-H_{\rm{max}}^{\overline{\varepsilon}} (\mathbf{X}_1|\mathbf{X}_1')\\
\geq& n_{Z_1}^{\rm{L}}\left(1-H_2(e_{Z_1}^{\rm{ph,U}})\right),
\end{aligned}
\end{equation} 
where $n_{Z_1}^{\rm{L}}$ is the lower bound of the length of $\mathbf{Z_1}$ with a failure probability $\varepsilon_{1}$, $e_{Z_1}^{\rm{ph,U}}$ is the upper bound of the single-photon phase error rate of $n_{Z_1}^{\rm{L}}$ with a failure probability $\varepsilon_e$. And $h(x)=-x\log_2 x -(1-x) \log_2 (1-x)$ is the binary Shannon entropy function. Actually, $e_{Z_1}^{\rm{ph,U}}$ can be estimated by the observed values. And the decoy-state method which is employed to estimated $n_{Z_1}^{\rm{L}}$ and $e_{Z_1}^{\rm{ph,U}}$ is shown in Supplementary Note C. 

If we choose 
\begin{equation}
\begin{aligned}
\varepsilon_{\rm{sec}}=2(\hat{\varepsilon}+2\overline{\varepsilon})+\varepsilon_{\rm{PA}},
\end{aligned}
\end{equation}
where $\overline{\varepsilon}=\sqrt{\varepsilon_e+\varepsilon_{1}}$, $\varepsilon_{\rm{PA}}$ is the failure probability of privacy amplification, we can get one of the main results of our paper by combining the equation behind,
\begin{equation}
\begin{aligned}
l_o \leq& n_{Z_1}^{\rm{L}}\left[ 1-h(e_{Z_1}^{\rm{ph,U}})\right]-{\rm{\lambda_{EC}}}\\
&-\log_2\frac{2}{\varepsilon_{\rm{cor}}}-2\log_2 \frac{1}{\sqrt{2}\hat{\varepsilon}\varepsilon_{\rm{PA}}}, 
\end{aligned}
\end{equation}
where ${\rm{\lambda_{EC}}}=f n_Z h (E_Z)$ is the information revealed in the error correction step, $f$ is the error correction efficiency which is related to the specific error correction scheme, $n_{Z}$ is the length of the raw key and $E_Z$ is the bit error rate between strings $\mathbf{Z}$ and $\mathbf{Z}'$. Moreover, the security coefficient of the whole protocol is $\varepsilon_{\rm{tol}}=\varepsilon_{\rm{cor}}+\varepsilon_{\rm{sec}}$, where $\varepsilon_{\rm{sec}}= 2(\hat{\varepsilon}+2\sqrt{\varepsilon_e+\varepsilon_{1}})+\varepsilon_{\rm{PA}}$. $\hat{\varepsilon}$ is the coefficient while using the chain rules, and $\varepsilon_{\rm{PA}}$ is the failure probability of privacy amplification.

\begin{widetext}
\section*{Supplemental Note A: Security analysis} \label{App1_Security analysis}
\subsection*{Outline of the proof}
For ease of understanding, we show the outline of our proof first, which can show the core of this analysis.
Let's assume $\rho_{AB}$ is a two-particle system, where $A$ and $B$ are controlled by Alice and Bob respectively. $A$ is spanned by two-qubit states 
$|01\rangle_A$ and $|01\rangle_A$. $B$ is defined similarly. Let's further define unitary operation $U^{\delta_a}_A |01\rangle_A = |01\rangle_A$, $U^{\delta_a}_A |10\rangle_A = e^{i \delta_a} |10\rangle_A$. $U^{\delta_b}_B$ is for qubit $B$ and defined analogously with $U^{\delta_a}_A$. Then we consider a protocol, in which Alice and Bob make Z-basis measurement on A and B to form key bit $Z_A$ and $Z_B$ after Alice and Bob applying $U^{\delta_a}_A$ and $U^{\delta_b}_B$ with probability mass function $p(\delta_a, \delta_b)$. This is equivalent to saying that $\rho'_{AB}= \int_{\delta_a}\int_{\delta_b}{\rm{d}\delta_a \rm{d}\delta_b} p(\delta_a,\delta_b)  U^{\delta_b}_{B} U^{\delta_a}_{A} \rho_{AB}  U^{-\delta_b}_{B} U^{-\delta_a}_{A}$ are considered as their system. To evaluate the conditional entropy $H(Z_A | E)$ in which $E$ is ancilla of Eve, it's of course that we can consider $Z_A$ is the outcome of Z-basis measurement on $\rho_{AB}$ rather than $\rho'_{AB}$. This is because the phases $\delta_a$ and $\delta_b$ have no physical effects on the $Z_A$ and $E$, the details are shown in the specific proof. In this view, we can just consider $\rho_{AB}$, and ignore the phases $\delta_a$ and $\delta_b$ if we only want to characterize $H(Z_A | E)$. Noting that the conjugate $X$-basis measurement is also made on $\rho_{AB}$ now. It's well known that $H(Z_A|E)$ can be bounded by $1 - H(X_A|X_B)$, where $X_A$ and $X_B$ is the outcome of X-basis ($\frac{|0\rangle+|1\rangle}{\sqrt{2}}$, $\frac{|0\rangle-|1\rangle}{\sqrt{2}}$) measurement on $\rho_{AB}$ and the error rate between $X_A$ and $X_B$ is called phase error rate.

Nevertheless, we can go further by following the logic in last paragraph. It's not restricted to assume $Z_A$ is measured on $\rho''_{AB}= \int_{\delta_a} \int_{\delta_b} q(\delta_a,\delta_b) U^{\delta_b}_{B} U^{\delta_a}_{A} \rho_{AB}  U^{-\delta_b}_{B} U^{-\delta_a}_{A}$ rather than ${\rho_{AB}}$, and phase error rate is of course defined to $\rho''_{AB}$ accordingly. Here, $q(\delta_a,\delta_b)$ is another probability mass function, which we can define to make it related to an experimentally observable value. The key point here is that $q(\delta_a,\delta_b)$ can be adjusted as we wish, since any $q(\delta_a,\delta_b)$ makes no difference for the Z-basis measurement. The freedom of defining $q(\delta_a,\delta_b)$ can help us list variant expressions of phase error rate, and then find appropriate one in some particular protocol.

Further, let's apply above considerations in MP-QKD protocol. $\rho_{AA'BB'}$ is a compound system for two adjacent rounds with successful detections i-th and j-th. $A$ is spanned by two-qubit states $|00\rangle_A$, $|01\rangle_A$, $|10\rangle_A$ and $|11\rangle_A$. $B$ is defined similarly. $A'$ and $B'$ are quantum registers recording the phases $\delta_a$ and $\delta_b$. Alice and Bob first measure $AB$ to decide if they are both in the Hilbert space spanned by $|01\rangle$ and $|10\rangle$. If so, this pair is a Z-pair and Z-basis measurement is followed to form key bit. We are interested in the phase error rate of a Z-pair, which are defined by the outcome of hypothetical X-basis ($\frac{|0\rangle+|1\rangle}{\sqrt{2}}$, $\frac{|0\rangle-|1\rangle}{\sqrt{2}}$) measurement on a Z-pair. Noting that following the logic in last paragraphs, we have the freedom to redefine the distribution $q(\delta_a,\delta_b)$ of $|\delta_a\rangle_{A'}$ and $|\delta_b\rangle_{B'}$ in $\rho_{AA'BB'}$ to find some appropriate expression of phase error rate. If Alice and Bob find $AB$ are both $|11\rangle$, then also $|\delta_a-\delta_b|<\Delta$ or $|\delta_a-\delta_b|>\pi-\Delta$ holds by measuring $A'$ and $B'$, this pair is an X-pair. Then Alice and Bob will calculate the bit error rate of this X-pair following some instructions. We proved that: 1. the ratio of probabilities for a pair happening to be a Z-pair or X-pair is fixed and cannot be affected by Eve; 2. By setting $q(\delta_a,\delta_b)$ satisfying $|\delta_a-\delta_b|<\Delta$ or $|\delta_a-\delta_b|>\pi-\Delta$. the phase error rate of a Z-pair just equals to the error rate of an X-pair. The two points guarantee that for any Z-pair, we have a fixed probability to test its phase error rate. The phase error rate is not directly measurable on $AB$, but it must equal to the bit error rate by measuring $A'B'$.

\subsection*{Original protocol}
In this section, we analyze the security of the original MP-QKD protocol. For simplicity, we omit the decoy state here, Alice and Bob prepare quantum states $\ket{0}$ and $\ket{\sqrt{\mu}e^{{\rm{i}}\theta}}$ with the probability $p_0$ and $p_1$ respectively. The phase $\theta$ of the quantum states are uniformly randomly chosen from $[0,2\pi)$. In the entanglement-based protocol, Alice and Bob prepare entangled states in each round,
\begin{equation}
\begin{aligned}
\ket{\Psi}_{AA'a}=\sqrt{p_0} \ket{0}_A \int_0^{2\pi} \frac{{\rm{d}}\theta}{\sqrt{2\pi}}  \ket{\theta}_{A'} \ket{0}_a +\sqrt{p_1} \ket{1}_A \int_0^{2\pi} \frac{{\rm{d}}\theta}{\sqrt{2\pi}} \ket{\theta}_{A'} \ket{\sqrt{\mu}e^{{\rm{i}}\theta}}_a,\\
\ket{\Psi}_{BB'b}=\sqrt{p_0} \ket{0}_B \int_0^{2\pi} \frac{{\rm{d}}\theta}{\sqrt{2\pi}} \ket{\theta}_{B'} \ket{0}_b +\sqrt{p_1} \ket{1}_B \int_0^{2\pi} \frac{{\rm{d}}\theta}{\sqrt{2\pi}} \ket{\theta}_{B'} \ket{\sqrt{\mu}e^{{\rm{i}}\theta}}_b,
\end{aligned}
\end{equation}
where $A(B)$ and $A'(B')$ are auxiliary quantum systems that are employed to store the intensity and random phase information, respectively. For the different values of $\theta$, the quantum states $|\theta\rangle$ are orthogonal, $\langle\theta|\theta'\rangle=\delta(\theta-\theta')$. Here we assume that the $i$-th round and the $j$-th round are two adjacent events with successful detection, which are recorded as $P^{i,j}$. The joint state of Alice can be written as
\begin{equation}
\begin{aligned}
\ket{\Psi}^{i,j}_{AA'a}&= p_0 |00\rangle_{A_iA_j} \int_0^{2\pi} \frac{{\rm{d}}\theta^i_a}{\sqrt{2\pi}} |\theta^i_a\rangle_{A'_i}|0\rangle_{a_i} \int_0^{2\pi} \frac{{\rm{d}}\theta^j_a}{\sqrt{2\pi}} |\theta^j_a\rangle_{A'_j}|0\rangle_{a_j}\\
&+\sqrt{p_0p_1}|01\rangle_{A_iA_j} \int_0^{2\pi} \frac{{\rm{d}}\theta^i_a}{\sqrt{2\pi}} |\theta^i_a\rangle_{A'_i}|0\rangle_{a_i} \int_0^{2\pi} \frac{{\rm{d}}\theta^j_a}{\sqrt{2\pi}} |\theta^j_a\rangle_{A'_j} |\sqrt{\mu}e^{{\rm{i}}\theta^j_a}\rangle_{a_j}\\
&+\sqrt{p_0p_1}|10\rangle_{A_iA_j} \int_0^{2\pi} \frac{{\rm{d}}\theta^i_a}{\sqrt{2\pi}} |\theta^i_a\rangle_{A'_i} |\sqrt{\mu}e^{{\rm{i}}\theta^i_a}\rangle_{a_i} \int_0^{2\pi} \frac{{\rm{d}}\theta^j_a}{\sqrt{2\pi}} |\theta^j_a\rangle_{A'_j}|0\rangle_{a_j}\\
&+p_1|11\rangle_{A_iA_j} \int_0^{2\pi} \frac{{\rm{d}}\theta^i_a}{\sqrt{2\pi}} |\theta^i_a\rangle_{A'_i} |\sqrt{\mu} e^{{\rm{i}}\theta^i_a}\rangle_{a_i} \int_0^{2\pi} \frac{{\rm{d}}\theta^j_a}{\sqrt{2\pi}} |\theta^j_a\rangle_{A'_j} |\sqrt{\mu}e^{{\rm{i}}\theta^j_a}\rangle_{a_j}.
\end{aligned}
\end{equation}
The joint state of Bob is given in a similar way.  When $A_iA_j$ collapses into the subspace spanned by $|01\rangle_{A_iA_j}$ and $|10\rangle_{A_iA_j}$, Alice is sure that $Z$-basis is chosen. $|11\rangle_{A_iA_j}$ means this pair is $X$-basis preparation.

In the $Z$-basis $P^{i,j}_{a(b)}$, the intensities are $\{0,\mu\}$ or $\{\mu,0\}$. Here we define $\theta^j_{a(b)}:=\theta^i_{a(b)}+\delta_{a(b)}$, the $Z$-basis preparation of Alice's joint state can be written as
\begin{equation}
\begin{aligned}
|\Psi_Z\rangle_{AA'a}^{i,j}=&\sqrt{p_0p_1}|01\rangle_{A_iA_j} \int_0^{2\pi} \frac{{\rm{d}}\theta^i_a}{\sqrt{2\pi}} \int_{-\pi}^{\pi} \frac{{\rm{d}}\delta_a}{\sqrt{2\pi}} |\theta^i_a\rangle_{A'_i}|0\rangle_{a_i}  |\delta_a\rangle_{A'_j} |\sqrt{\mu}e^{{\rm{i}}(\theta^i_a+\delta_a)}\rangle_{a_j}\\
&+\sqrt{p_0p_1}|10\rangle_{A_iA_j} \int_0^{2\pi} \frac{{\rm{d}}\theta^i_a}{\sqrt{2\pi}} \int_{-\pi}^{\pi} \frac{{\rm{d}}\delta_a}{\sqrt{2\pi}} |\theta^i_a\rangle_{A'_i} |\sqrt{\mu}e^{{\rm{i}}\theta^i_a}\rangle_{a_i}  |\delta_a\rangle_{A'_j}|0\rangle_{a_j}\\
=&\sqrt{p_0p_1} \int_0^{2\pi} \frac{{\rm{d}}\theta^i_a}{\sqrt{2\pi}} \int_{-\pi}^{\pi} \frac{{\rm{d}}\delta_a}{\sqrt{2\pi}} |\theta^i_a\rangle_{A'_i} |\delta_a\rangle_{A'_j} \\
& \sum_{n=0}^\infty \sqrt{p_{n|\mu}} e^{{\rm{i}}n\theta^i_a}(e^{{\rm{i}}n\delta_a}|01\rangle_{A_iA_j}|0\rangle_{a_i} |n\rangle_{a_j} + |10\rangle_{A_iA_j}|n\rangle_{a_i} |0\rangle_{a_j}),
\end{aligned}
\end{equation}
where $n$ is the number of the photon, and $p_{n|\mu}= \frac{e^{-\mu} \mu^n}{n!}$ is the probability of the Poisson distribution.

First, we trace the auxiliary $A'_i$, the density matrix of Alice's $Z$-basis preparation can be written as
\begin{equation}
\begin{aligned}
\rho_{a,Z}^{i,j}=&\text{Tr}_{A'_i} (|\Psi_Z\rangle \langle\Psi_Z|^{i,j}_{AA'a})\\
=&p_0 p_1 \int_{-\pi}^{\pi} \frac{{\rm{d}}\delta_a}{\sqrt{2\pi}} \int_{-\pi}^{\pi} \frac{{\rm{d}}\delta'_a}{\sqrt{2\pi}} |\delta_a\rangle\langle\delta'_a|_{A'_j} \sum_{n=0}^{\infty} p_{n|\mu} (e^{{\rm{i}}n\delta_a}|01\rangle_{A_iA_j}|0\rangle_{a_i} |n\rangle_{a_j} + |10\rangle_{A_iA_j}|n\rangle_{a_i} |0\rangle_{a_j}) \\
&\times(e^{{\rm{i}}n\delta'_a}\langle01|_{A_iA_j}\langle0|_{a_i} \langle n|_{a_j} + \langle10|_{A_iA_j}\langle n|_{a_i} \langle 0|_{a_j})\\
=&2 p_0 p_1 p_{1|\mu} \bm{\mathcal{P}} \left[ \int_{-\pi}^{\pi} \frac{{\rm{d}}\delta_a}{\sqrt{2\pi}} |\delta_a\rangle \frac{ |01\rangle_{A_iA_j}|01\rangle_{a_ia_j} + e^{-{\rm{i}}\delta_a}|10\rangle_{A_iA_j}|10\rangle_{a_ia_j} }{\sqrt{2}}\right]+{\rm{Part_{\rm{zm}}}},
\end{aligned}
\end{equation} 
where $\bm{\mathcal{P}}[|\cdot\rangle]=|\cdot\rangle\langle\cdot|$, and ${\rm{Part_{\rm{zm}}}}$ is the mixture components of the zero-photon and multiple-photon in $a_ia_j$. Because the system $A'_j$ is measured and announced in the subsequent steps, the single-photon component of the density matrix can be simplified as a classical-quantum state,  i.e.  
\begin{equation}
\begin{aligned}
\rho_{a,Z_1}^{i,j}=2 p_0 p_1 p_{1|\mu} \int_{-\pi}^{\pi} \frac{{\rm{d}}\delta_a}{2\pi} |\delta_a\rangle\langle\delta_a|_{A_j'} \bm{\mathcal{P}}\left[\frac{|01\rangle_{A_iA_j}|01\rangle_{a_ia_j} + e^{-i\delta_a}|10\rangle_{A_iA_j}|10\rangle_{a_ia_j}}{\sqrt{2}}\right].
\end{aligned}
\end{equation}

Actually, when Alice takes the $\hat{Z}_A$ measurement on $A_iA_j$, whose eigenstates are $|01\rangle_{A_iA_j}$ and $|10\rangle_{A_iA_j}$, to the density matrix $\rho_{a,Z_1}^{i,j}$, the phase $\delta_a$ plays no roles in the results of the measurement and Eve's potential system. Therefore, it's not restrictive to introduce a unitary operation $U_A^{\delta_a}$ to the auxiliaries $A_i,A_j$ before the $\hat{Z}_A$ measurement, defined by $U_A^{\delta_a}|01\rangle_{A_iA_j}=|01\rangle_{A_iA_j}$, $U_A^{\delta_a}|10\rangle_{A_iA_j}=e^{i\delta_a}|10\rangle_{A_iA_j}$. This operation can keep the specific form of send states $a_i$ and $a_j$. The density matrix can be simplified as 
\begin{equation}
\begin{aligned}
\rho_{a,Z_1}^{i,j}=2 p_0 p_1 p_{1|\mu} \int_{-\pi}^{\pi} \frac{{\rm{d}}\delta_a}{2\pi} |\delta_a\rangle\langle\delta_a|_{A_j'} \bm{\mathcal{P}}\left[\frac{|01\rangle_{A_iA_j}|01\rangle_{a_ia_j} + |10\rangle_{A_iA_j}|10\rangle_{a_ia_j}}{\sqrt{2}}\right].
\end{aligned}
\end{equation}

Then we consider the single-photon component of the joint density matrix of Alice and Bob, given by
\begin{equation}
\begin{aligned}
\rho_{Z_1}^{i,j}=& 4 p_0^2 p_1^2 p_{1|\mu}^2  \int_{-\pi}^{\pi} \frac{{\rm{d}}\delta_a}{2\pi} \int_{-\pi}^{\pi} \frac{{\rm{d}}\delta_b}{2\pi} |\delta_a\delta_b\rangle\langle \delta_a\delta_b|_{A_j'B_j'} \\
& \otimes \hat{M}_E \bm{\mathcal{P}}\left[ \frac{|01\rangle_{A_iA_j}|01\rangle_{a_ia_j} + |10\rangle_{A_iA_j}|10\rangle_{a_ia_j}}{\sqrt{2}} \right] \otimes \bm{\mathcal{P}}\left[ \frac{|01\rangle_{B_iB_j}|01\rangle_{b_ib_j} + |10\rangle_{B_iB_j}|10\rangle_{b_ib_j}}{\sqrt{2}} \right] \hat{M}^\dagger_E,
\end{aligned}
\end{equation}
where  $\hat{M}_E$ is an element in a set of POVM measurements made by Eve, with outputs $C^i=C^j=1$. It should be noted that $\hat{M}_E$ operates on the systems $a_i$, $a_j$, $b_i$, and $b_j$ which are sent to Charlie. Here we just care about the reduced density matrix of $i$-th and $j$-th round. For different $i$ and $j$, $\hat{M}_E$ might be different and relate to other rounds, $\hat{M}_E^{i,j}$ is a more suitable symbol. For simplicity, we omitted the superscript here. In the subsequent proof, we can see that the specific form of $\hat{M}_E^{i,j}$ is needless. For the $Z$-basis preparation, the systems $A_iA_ja_ia_j$ and $B_iB_jb_ib_j$ are independent of the values $\delta_a$ and $\delta_b$, which are never announced. So, we can trace the auxiliaries $A'_jB'_j$ for simplicity. We have
\begin{equation}
\begin{aligned}
\tilde{\rho}_{Z_1}^{i,j}=& \Tr_{A'_jB'_j}\left(\rho_{Z_1}^{i,j}\right)\\
=& 4 p_0^2 p_1^2 p_{1|\mu}^2 \hat{M}_E \bm{\mathcal{P}}\left[ \frac{|01\rangle_{A_iA_j}|01\rangle_{a_ia_j} + |10\rangle_{A_iA_j}|10\rangle_{a_ia_j}}{\sqrt{2}} \right] \otimes \bm{\mathcal{P}}\left[ \frac{|01\rangle_{B_iB_j}|01\rangle_{b_ib_j} + |10\rangle_{B_iB_j}|10\rangle_{b_ib_j}}{\sqrt{2}} \right] \hat{M}^\dagger_E.
\end{aligned}
\end{equation}
For obtaining the single-photon phase error rate of the $Z$-basis preparation, we define two measurements $\hat{X}_A$ and $\hat{X}_B$, whose eigenstates are 
\begin{equation}
\begin{aligned}
|X_A^{+(-)}\rangle_{A_iA_j}:=\frac{ |01\rangle_{A_iA_j} \pm |10\rangle_{A_iA_j}}{\sqrt{2}},
|X_B^{+(-)}\rangle_{B_iB_j}:=\frac{ |01\rangle_{B_iB_j} \pm |10\rangle_{B_iB_j}}{\sqrt{2}},
\end{aligned}
\end{equation}
respectively. The joint density matrix can be revised by these eigenstates, 
\begin{equation}
\begin{aligned}
\tilde{\rho}_{Z_1}^{i,j}=& 4 p_0^2 p_1^2 p_{1|\mu}^2 \hat{M}_E \bm{\mathcal{P}}\left[ \frac{1}{\sqrt{2}}\left(|X_{A}^{+}\rangle_{A_iA_j} \frac{ |01\rangle_{a_i a_j}+|10\rangle_{a_ia_j}}{\sqrt{2}} + |X_{A}^{-}\rangle_{A_iA_j} \frac{|01\rangle_{a_i a_j}-|10\rangle_{a_ia_j}}{\sqrt{2}}\right) \right] \\
& \otimes \bm{\mathcal{P}}\left[ \frac{1}{\sqrt{2}} \left(|X_{B}^{+}\rangle_{B_iB_j} \frac{|01\rangle_{b_i b_j}+|10\rangle_{b_ib_j}}{\sqrt{2}} + |X_{B}^{-}\rangle_{B_iB_j} \frac{|01\rangle_{b_i b_j}-|10\rangle_{b_ib_j}}{\sqrt{2}}\right) \right] \hat{M}^\dagger_E  .
\end{aligned}
\end{equation}
The probability of projection into $|X_{A}^{+}\rangle_{A_iA_j}|X_{B}^{+}\rangle_{B_iB_j}$ can be given by
\begin{equation}
\begin{aligned}
{\rm{Pr}}(|X_{A}^{+}\rangle_{A_iA_j}|X_{B}^{+}\rangle_{B_iB_j})=&\Tr{\langle X_{A}^{+}|_{A_iA_j} \langle X_{B}^{+}|_{B_iB_j} \tilde{\rho}_{Z_1}^{i,j} |X_{A}^{+}\rangle_{A_iA_j}|X_{B}^{+}\rangle_{B_iB_j}}\\
=& p_0^2 p_1^2 p_{1|\mu}^2 \Tr \left\{ \hat{M}_E \bm{\mathcal{P}} \left[  \frac{ |01\rangle_{a_i a_j}+|10\rangle_{a_ia_j}}{\sqrt{2}} \otimes \frac{|01\rangle_{b_i b_j}+|10\rangle_{b_ib_j}}{\sqrt{2}} \right] \hat{M}^\dagger_E \right\}.
\end{aligned}
\end{equation}
Similarly,
\begin{equation}
\begin{aligned}
{\rm{Pr}}(|X_{A}^{+}\rangle_{A_iA_j}|X_{B}^{-}\rangle_{B_iB_j})=& p_0^2 p_1^2 p_{1|\mu}^2 \Tr \left\{ \hat{M}_E \bm{\mathcal{P}} \left[  \frac{ |01\rangle_{a_i a_j}+|10\rangle_{a_ia_j}}{\sqrt{2}} \otimes \frac{|01\rangle_{b_i b_j}-|10\rangle_{b_ib_j}}{\sqrt{2}} \right] \hat{M}^\dagger_E \right\},\\
{\rm{Pr}}(|X_{A}^{-}\rangle_{A_iA_j}|X_{B}^{+}\rangle_{B_iB_j})=& p_0^2 p_1^2 p_{1|\mu}^2 \Tr \left\{ \hat{M}_E \bm{\mathcal{P}} \left[  \frac{ |01\rangle_{a_i a_j}-|10\rangle_{a_ia_j}}{\sqrt{2}} \otimes \frac{|01\rangle_{b_i b_j}+|10\rangle_{b_ib_j}}{\sqrt{2}} \right] \hat{M}^\dagger_E \right\},\\
{\rm{Pr}}(|X_{A}^{-}\rangle_{A_iA_j}X_{B}^{-}\rangle_{B_iB_j})=& p_0^2 p_1^2 p_{1|\mu}^2 \Tr \left\{ \hat{M}_E \bm{\mathcal{P}} \left[  \frac{ |01\rangle_{a_i a_j}-|10\rangle_{a_ia_j}}{\sqrt{2}} \otimes \frac{|01\rangle_{b_i b_j}-|10\rangle_{b_ib_j}}{\sqrt{2}} \right] \hat{M}^\dagger_E \right\}.
\end{aligned}
\end{equation}
Moreover,
\begin{equation}
\begin{aligned}
&\left[\Pr(|X_{A}^{+}\rangle|X_{B}^{+}\rangle)+
\Pr(|X_{A}^{+}\rangle|X_{B}^{-}\rangle)+
\Pr(|X_{A}^{-}\rangle|X_{B}^{+}\rangle)+
\Pr(|X_{A}^{-}\rangle|X_{B}^{-}\rangle)\right]_{A_iA_j,B_iB_j}\\
=& p_0^2 p_1^2 p_{1|\mu}^2 \Tr \left\{\hat{M}_E \left\{ \bm{\mathcal{P}}\left[ \frac{ |01\rangle_{a_i a_j}+|10\rangle_{a_ia_j}}{\sqrt{2}} \right] + \bm{\mathcal{P}}\left[\frac{|01\rangle_{a_i a_j}-|10\rangle_{a_ia_j}}{\sqrt{2}} \right] \right\} \right.\\
& \left. \otimes \left\{ \bm{\mathcal{P}} \left[ \frac{ |01\rangle_{b_i b_j}+|10\rangle_{b_ib_j}}{\sqrt{2}} \right] + \bm{\mathcal{P}}\left[\frac{|01\rangle_{b_i b_j}-|10\rangle_{b_ib_j}}{\sqrt{2}} \right] \right\} \hat{M}^\dagger_E \right\}\\
=&\Tr \tilde\rho_{Z_1}^{i,j}.
\end{aligned}
\end{equation}

In this way, the single-photon phase error rate of $Z$-basis preparation can be obtained by
\begin{equation}
\label{ephz1}
\begin{aligned}
e_{Z_1}^{\rm{ph}}(\tilde{\rho}_{Z_1}^{i,j})=&
\frac{{\rm{Pr}}(|X_{A}^{+}\rangle_{A_iA_j}|X_{B}^{-}\rangle_{B_iB_j})+
{\rm{Pr}}(|X_{A}^{-}\rangle_{A_iA_j}|X_{B}^{+}\rangle_{B_iB_j})}
{\left[\Pr(|X_{A}^{+}\rangle|X_{B}^{+}\rangle)+
\Pr(|X_{A}^{+}\rangle|X_{B}^{-}\rangle)+
\Pr(|X_{A}^{-}\rangle|X_{B}^{+}\rangle)+
\Pr(|X_{A}^{-}\rangle|X_{B}^{-}\rangle)\right]_{A_iA_j,B_iB_j}}\\
=&\frac{p_0^2p_1^2p_{1|\mu}^2}{\Tr \tilde\rho_{Z_1}^{i,j}}  \Tr \left\{ \hat{M}_E \bm{\mathcal{P}} \left[  \frac{ |01\rangle_{a_i a_j}+|10\rangle_{a_ia_j}}{\sqrt{2}} \otimes \frac{|01\rangle_{b_i b_j}-|10\rangle_{b_ib_j}}{\sqrt{2}} \right] \hat{M}^\dagger_E \right\} \\
+&\frac{p_0^2p_1^2p_{1|\mu}^2}{\Tr \tilde\rho_{Z_1}^{i,j}}  \Tr \left\{ \hat{M}_E \bm{\mathcal{P}} \left[  \frac{ |01\rangle_{a_i a_j}-|10\rangle_{a_ia_j}}{\sqrt{2}} \otimes \frac{|01\rangle_{b_i b_j}+|10\rangle_{b_ib_j}}{\sqrt{2}} \right] \hat{M}^\dagger_E \right\} .\\
\end{aligned}
\end{equation}

The security proof focuses on how to estimate this phase error rate. To solve this problem, we resort to $X$-pair preparation. In $X$-basis $P_{a(b)}^{i,j}$, the intensities is $\{\mu,\mu\}$. For the case that Alice prepares $X$-basis, the joint states of Alice can be written as 
\begin{equation}
\begin{aligned}
|\Psi_X\rangle_{AA'a}^{i,j}=p_1 |11\rangle_{A_iA_j}\int_{0}^{2\pi} \frac{{\rm{d}}\theta^i_a}{\sqrt{2\pi}} \int_{-\pi}^{\pi} \frac{{\rm{d}}\delta_a}{\sqrt{2\pi}} |\theta_a^i\rangle_{A'_i} |\delta_a\rangle_{A'_j} |\sqrt{\mu} e^{{\rm{i}}\theta_a^i}\rangle_{a_i}  |\sqrt{\mu} e^{{\rm{i}}(\theta_a^i+\delta_a)}\rangle_{a_j}. 
\end{aligned}
\end{equation}
Just like the operation we did to the $Z$-basis preparation, the auxiliary $A'_i$ is traced first, the density matrix of Alice's $X$-basis preparation can be written as
\begin{equation}
\begin{aligned}
\rho_{a,X}^{i,j}=&\Tr_{A'_i} \left(|\Psi_X\rangle \langle\Psi_X|^{i,j}_{AA'a}\right)\\
=&p_1^2 |11\rangle\langle11|_{A_iA_j}\int_{-\pi}^{\pi}\frac{{\rm{d}}\delta_a}{\sqrt{2\pi}} \int_{-\pi}^{\pi}\frac{{\rm{d}}\delta'_a}{\sqrt{2\pi}} |\delta_a\rangle\langle \delta'_a|_{A'_j}  \sum_{n_1+n_2=m_1+m_2}\sqrt{p_{n_1|\mu}p_{n_2|\mu}p_{m_1|\mu}p_{m_2|\mu}}e^{{\rm{i}}{(n_2\delta_a-m_2\delta'_a)}}|n_1n_2 \rangle_{a_ia_j}\langle m_1m_2|\\
=&p_1^2 |11\rangle\langle11|_{A_iA_j}\int_{-\pi}^{\pi}\frac{{\rm{d}}\delta_a}{\sqrt{2\pi}} \int_{-\pi}^{\pi}\frac{{\rm{d}}\delta'_a}{\sqrt{2\pi}} |\delta_a\rangle\langle \delta'_a|_{A'_j}  \{p_{1|\mu}p_{0|\mu}(|01\rangle_{a_ia_j}+e^{-{\rm{i}}\delta_a}|10\rangle_{a_ia_j})(\langle 01|_{a_ia_j}+e^{{\rm{i}}\delta'_a}\langle 10|_{a_ia_j})\}+ {\rm{Part_{\rm{zm}}}}\\
=&p_1^2 p_{1|2\mu} |11\rangle\langle11|_{A_iA_j} \bm{\mathcal{P}}\left[ \int_{-\pi}^{\pi} \frac{{\rm{d}}\delta_a}{\sqrt{2\pi}} |\delta_a\rangle_{A'_j} \frac{|01\rangle_{a_ia_j}+e^{-{\rm{i}}\delta_a}|10\rangle_{a_ia_j}}{\sqrt{2}}  \right]+{\rm{Part_{\rm{zm}}}}\\
=& p_1^2 p_{1|2\mu} |11\rangle\langle11|_{A_iA_j}  \bm{\mathcal{P}}\left[ \int_{0}^{\pi} \frac{{\rm{d}}\delta_a}{\sqrt{2\pi}}\left(|\delta_a^+\rangle_{A'_j}\frac{|01\rangle_{a_ia_j}+e^{-{\rm{i}}\delta_a}|10\rangle_{a_ia_j}}{\sqrt{2}}+|\delta_a^-\rangle_{A'_j}\frac{|01\rangle_{a_ia_j}-e^{-{\rm{i}}\delta_a}|10\rangle_{a_ia_j}}{\sqrt{2}}\right)\right]\\
&+{\rm{Part_{\rm{zm}}}},
\end{aligned}
\end{equation}
where $\delta_a^+=\delta_a$ and $\delta_a^-=\delta_a +\pi$. Indeed $|\delta_a^+\rangle$ $(|\delta_a^-\rangle)$ corresponds to logical bit $0$ $(1)$ with $\delta_a$ in $X$-basis.
For example, in the original protocol, if Alice finds the phase difference between the $i$-th and $j$-th pulse is $8\pi/7$, this is equivalent to that his measurement result of $A'_j$ is $|\delta_a^-\rangle$ (logical bit $1$) with $\delta_a=\pi/7$ in the entanglement-based protocol.

The density matrix of Bob preparing $X$-basis is the same. Then we consider the single-photon component of the joint density matrix of Alice and Bob,
\begin{equation}
\begin{aligned}
\rho_{X_1}^{i,j}=& p_1^4 p_{1|2\mu}^2 \bm{\mathcal{P}}[|11\rangle_{A_iA_j}|11\rangle_{B_iB_j}] \\
&\times  \hat{M}_E  \bm{\mathcal{P}}\left[\int_{0}^{\pi} \frac{{\rm{d}}\delta_a}{\sqrt{2\pi}} \left(|\delta_a^+\rangle_{A'_j}\frac{|01\rangle_{a_ia_j}+e^{-{\rm{i}}\delta_a}|10\rangle_{a_ia_j}}{\sqrt{2}}+|\delta_a^-\rangle_{A'_j}\frac{|01\rangle_{a_ia_j}-e^{-{\rm{i}}\delta_a}|10\rangle_{a_ia_j}}{\sqrt{2}}\right)\right]\\
&\otimes  \bm{\mathcal{P}}\left[ \int_{0}^{\pi} \frac{{\rm{d}}\delta_b}{\sqrt{2\pi}} \left(|\delta_b^+\rangle_{B'_j}\frac{|01\rangle_{b_ib_j}+e^{-{\rm{i}}\delta_b}|10\rangle_{b_ib_j}}{\sqrt{2}}+|\delta_b^-\rangle_{B'_j}\frac{|01\rangle_{b_ib_j}-e^{-{\rm{i}}\delta_b}|10\rangle_{b_ib_j}}{\sqrt{2}}\right)\right] \hat{M}^\dagger_E.
\end{aligned}
\end{equation}

Here we define two measurements $\hat{X}^{\delta_a}_{A'}$ and $\hat{X}^{\delta_b}_{B'}$, whose eigenstates are $|\delta_a^ \pm \rangle_{A'_j}$ and $|\delta_b^ \pm\rangle_{B'_j}$, respectively. The preparation of quantum states can be regarded as Alice and Bob taking the above measurements to the auxiliaries $A'_j$ and $B'_j$. The joint probability that $C^i=C^j=1$, Alice and Bob prepare the single-photon component of X-pair and the measurement results are $|\delta_a^+\rangle_{A_j'}$ and $|\delta_b^+\rangle_{B_j'}$ is
\begin{equation}
\begin{aligned}
{\rm{Pr}}\left(|\delta_a^+\rangle_{A'_j}|\delta_b^+\rangle_{B'_j}\right)=&
\Tr{\langle \delta_a^{+}|_{A'_j} \langle \delta_b^{+}|_{B'_j} \rho_{X_1}^{i,j} |\delta_a^{+}\rangle_{A'_j}|\delta_b^{+}\rangle_{B'_j}}\\
=& \frac{p_1^4 p_{1|2\mu}^2 {\rm{d}}\delta_a {\rm{d}}\delta_b}{4 \pi^2}  \Tr \left\{ \hat{M}_E \bm{\mathcal{P}} \left[  \frac{( |01\rangle_{a_i a_j}+e^{-{\rm{i}}\delta_a} |10\rangle_{a_ia_j})}{\sqrt{2}} \otimes \frac{(|01\rangle_{b_i b_j}+e^{-{\rm{i}}\delta_b} |10\rangle_{b_ib_j})}{\sqrt{2}} \right] \hat{M}^\dagger_E \right\}.\\
\end{aligned}
\end{equation}
Similarly,
\begin{equation}
\begin{aligned}
{\rm{Pr}}\left(|\delta_a^+\rangle_{A'_j}|\delta_b^-\rangle_{B'_j}\right)=& \frac{p_1^4 p_{1|2\mu}^2 {\rm{d}}\delta_a {\rm{d}}\delta_b}{4 \pi^2}  \Tr \left\{ \hat{M}_E \bm{\mathcal{P}} \left[  \frac{( |01\rangle_{a_i a_j}+e^{-{\rm{i}}\delta_a} |10\rangle_{a_ia_j})}{\sqrt{2}} \otimes \frac{(|01\rangle_{b_i b_j}-e^{-{\rm{i}}\delta_b} |10\rangle_{b_ib_j})}{\sqrt{2}} \right] \hat{M}^\dagger_E \right\},\\
{\rm{Pr}}\left(|\delta_a^-\rangle_{A'_j}|\delta_b^+\rangle_{B'_j}\right)=& \frac{p_1^4 p_{1|2\mu}^2 {\rm{d}}\delta_a {\rm{d}}\delta_b}{4 \pi^2}  \Tr \left\{ \hat{M}_E \bm{\mathcal{P}} \left[  \frac{( |01\rangle_{a_i a_j}-e^{-{\rm{i}}\delta_a} |10\rangle_{a_ia_j})}{\sqrt{2}} \otimes \frac{(|01\rangle_{b_i b_j}+e^{-{\rm{i}}\delta_b} |10\rangle_{b_ib_j})}{\sqrt{2}} \right] \hat{M}^\dagger_E \right\},\\
{\rm{Pr}}\left(|\delta_a^-\rangle_{A'_j}|\delta_b^-\rangle_{B'_j}\right)=& \frac{p_1^4 p_{1|2\mu}^2 {\rm{d}}\delta_a {\rm{d}}\delta_b}{4 \pi^2}  \Tr \left\{ \hat{M}_E \bm{\mathcal{P}} \left[  \frac{( |01\rangle_{a_i a_j}-e^{-{\rm{i}}\delta_a} |10\rangle_{a_ia_j})}{\sqrt{2}} \otimes \frac{(|01\rangle_{b_i b_j}-e^{-{\rm{i}}\delta_b} |10\rangle_{b_ib_j})}{\sqrt{2}} \right] \hat{M}^\dagger_E \right\}.\\
\end{aligned}
\end{equation}

For obtaining single-photon bit error rate with $\delta_a$ and $\delta_b$, the sum of ${\rm{Pr}}\left(|\delta_a^\pm\rangle_{A'_j}|\delta_b^\pm\rangle_{B'_j}\right)$ is needed, which is given by
\begin{equation}
\begin{aligned}
{\rm{Pr}}(\delta_a\delta_b)=&{\rm{Pr}}\left(|\delta_a^+\rangle_{A'_j}|\delta_b^+\rangle_{B'_j}\right)+
{\rm{Pr}}\left(|\delta_a^+\rangle_{A'_j}|\delta_b^-\rangle_{B'_j}\right)+
{\rm{Pr}}\left(|\delta_a^-\rangle_{A'_j}|\delta_b^+\rangle_{B'_j}\right)+
{\rm{Pr}}\left(|\delta_a^-\rangle_{A'_j}|\delta_b^-\rangle_{B'_j}\right) \\ 
=& \frac{p_1^4 p_{1|2\mu}^2 {\rm{d}}\delta_a {\rm{d}}\delta_b}{4\pi^2} \Tr \left\{ \hat{M}_E \left\{ \bm{\mathcal{P}}\left[ \frac{|01\rangle_{a_ia_j}+e^{-{\rm{i}}\delta_a}|10\rangle_{a_ia_j}}{\sqrt{2}}\right] + \bm{\mathcal{P}}\left[ \frac{|01\rangle_{a_ia_j}-e^{-{\rm{i}}\delta_a}|10\rangle_{a_ia_j}}{\sqrt{2}} \right] \right\} \right.\\
& \otimes \left. \left\{ \bm{\mathcal{P}}\left[ \frac{|01\rangle_{b_ib_j}+e^{-{\rm{i}}\delta_b}|10\rangle_{b_ib_j}}{\sqrt{2}} \right] +  \bm{\mathcal{P}}\left[ \frac{|01\rangle_{b_ib_j}-e^{-{\rm{i}}\delta_b}|10\rangle_{b_ib_j}}{\sqrt{2}} \right] \right\} \hat{M}^\dagger_E \right\}\\
=& \frac{p_1^4 p_{1|2\mu}^2 {\rm{d}}\delta_a {\rm{d}}\delta_b}{4\pi^2} \Tr \left\{ \hat{M}_E \left( \frac{ |01 \rangle_{a_ia_j} \langle 01| }{2} + \frac{ |10 \rangle_{a_ia_j} \langle 10| }{2} \right) \right. \otimes \left. \left( \frac{ |01 \rangle_{b_ib_j} \langle 01| }{2} + \frac{ |10 \rangle_{b_ib_j} \langle 10| }{2} \right) \hat{M}^\dagger_E \right\}\\
=&\frac{p_1^4 p_{1|2\mu}^2 {\rm{d}}\delta_a {\rm{d}}\delta_b}{4\pi^2} \Tr \left\{ \hat{M}_E \left\{ \bm{\mathcal{P}}\left[ \frac{|01\rangle_{a_ia_j}+|10\rangle_{a_ia_j}}{\sqrt{2}}\right] + \bm{\mathcal{P}}\left[ \frac{|01\rangle_{a_ia_j}-|10\rangle_{a_ia_j}}{\sqrt{2}} \right] \right\} \right.\\
& \otimes \left. \left\{ \bm{\mathcal{P}}\left[ \frac{|01\rangle_{b_ib_j}+|10\rangle_{b_ib_j}}{\sqrt{2}} \right] + \bm{\mathcal{P}}\left[ \frac{|01\rangle_{b_ib_j}-|10\rangle_{b_ib_j}}{\sqrt{2}} \right] \right\} \hat{M}^\dagger_E \right\}.\\
\end{aligned}
\end{equation}
Moreover, 
\begin{equation}
\begin{aligned}
\Tr \rho_{X_1}^{i,j}=
&p_1^4 p_{1|2\mu}^2 \int_{0}^{\pi} \frac{{\rm{d}}\delta_a}{2\pi} \int_{0}^{\pi} \frac{{\rm{d}}\delta_b}{2\pi} 
\Tr \left\{ \hat{M}_E  \bm{\mathcal{P}}\left[ |\delta_a^+\rangle_{A'_j}\frac{|01\rangle_{a_ia_j}+e^{-{\rm{i}}\delta_a}|10\rangle_{a_ia_j}}{\sqrt{2}}+|\delta_a^-\rangle_{A'_j}\frac{|01\rangle_{a_ia_j}-e^{-{\rm{i}}\delta_a}|10\rangle_{a_ia_j}}{\sqrt{2}} \right] \right.\\
& \otimes \left. \bm{\mathcal{P}}\left[ |\delta_b^+\rangle_{B'_j}\frac{|01\rangle_{b_ib_j}+e^{-{\rm{i}}\delta_b}|10\rangle_{b_ib_j}}{\sqrt{2}}+|\delta_b^-\rangle_{B'_j}\frac{|01\rangle_{b_ib_j}-e^{-{\rm{i}}\delta_b}|10\rangle_{b_ib_j}}{\sqrt{2}} \right] \hat{M}^\dagger_E \right\}\\
=& p_1^4 p_{1|2\mu}^2 \int_{0}^{\pi} \frac{{\rm{d}}\delta_a}{2\pi} \int_{0}^{\pi} \frac{{\rm{d}}\delta_b}{2\pi} 
 \Tr \left\{ \hat{M}_E \left\{ \bm{\mathcal{P}}\left[ \frac{|01\rangle_{a_ia_j}+e^{-{\rm{i}}\delta_a}|10\rangle_{a_ia_j}}{\sqrt{2}}\right] + \bm{\mathcal{P}}\left[ \frac{|01\rangle_{a_ia_j}-e^{-{\rm{i}}\delta_a}|10\rangle_{a_ia_j}}{\sqrt{2}} \right] \right\} \right.\\
& \otimes \left. \left\{ \bm{\mathcal{P}}\left[ \frac{|01\rangle_{b_ib_j}+e^{-{\rm{i}}\delta_b}|10\rangle_{b_ib_j}}{\sqrt{2}} \right] + \bm{\mathcal{P}}\left[ \frac{|01\rangle_{b_ib_j}-e^{-{\rm{i}}\delta_b}|10\rangle_{b_ib_j}}{\sqrt{2}} \right] \right\} \hat{M}^\dagger_E \right\}\\
=& p_1^4 p_{1|2\mu}^2 \int_{0}^{\pi} \frac{{\rm{d}}\delta_a}{2\pi} \int_{0}^{\pi} \frac{{\rm{d}}\delta_b}{2\pi} \Tr \left\{ \hat{M}_E \left\{ \bm{\mathcal{P}}\left[ \frac{|01\rangle_{a_ia_j}+|10\rangle_{a_ia_j}}{\sqrt{2}}\right] + \bm{\mathcal{P}}\left[ \frac{|01\rangle_{a_ia_j}-|10\rangle_{a_ia_j}}{\sqrt{2}} \right] \right\} \right.\\
& \otimes \left. \left\{ \bm{\mathcal{P}}\left[ \frac{|01\rangle_{b_ib_j}+|10\rangle_{b_ib_j}}{\sqrt{2}} \right] + \bm{\mathcal{P}}\left[ \frac{|01\rangle_{b_ib_j}-|10\rangle_{b_ib_j}}{\sqrt{2}} \right] \right\} \hat{M}^\dagger_E \right\}\\
=&\frac{p_1^4 p_{1|2\mu}^2}{4} \Tr \left\{ \hat{M}_E \left\{ \bm{\mathcal{P}}\left[ \frac{|01\rangle_{a_ia_j}+|10\rangle_{a_ia_j}}{\sqrt{2}}\right] + \bm{\mathcal{P}}\left[ \frac{|01\rangle_{a_ia_j}-|10\rangle_{a_ia_j}}{\sqrt{2}} \right] \right\} \right.\\
& \otimes \left. \left\{ \bm{\mathcal{P}}\left[ \frac{|01\rangle_{b_ib_j}+|10\rangle_{b_ib_j}}{\sqrt{2}} \right] + \bm{\mathcal{P}}\left[ \frac{|01\rangle_{b_ib_j}-|10\rangle_{b_ib_j}}{\sqrt{2}} \right] \right\} \hat{M}^\dagger_E \right\}.\\
\end{aligned}
\end{equation}
Then it's clear that 
\begin{equation}
\Pr \left( \delta_a \delta_b \right)=\frac{{\rm{d}}\delta_a {\rm{d}}\delta_b}{\pi^2} \Tr \rho_{X_1}^{i,j},
\end{equation}
which means that Eve's attacks cannot depend on the phases $\delta_a$ and $\delta_b$, i.e. $\Pr \left( \delta_a \delta_b \right)$ does not depend on the phases $\delta_a$ and $\delta_b$.

Actually, in the key mapping step, Bob keeps his logical bit if $|\delta_a-\delta_b|\leq\Delta$, flips his logical bit if $|\delta_a-\delta_b|\geq\pi-\Delta$. And he keeps his logical bit if Charlie's clicks are (L,L) or (R,R), flips his logical bit if (L,R) or (R,L).
In the entanglement-based protocol, Bob operates his auxiliary $B'_j$ to implement this reverse.
For X-pair preparation, the single-photon bit error rate under the condition Alice and Bob obtained the results $\delta_a$ and $\delta_b$ ($\delta_{a(b)}\in[0,\pi)$, $|\delta_a-\delta_b|\leq\Delta$) is
\begin{equation}
\begin{aligned}
e_{X_1}^{\rm{bit}}(\delta_a,\delta_b)=& \frac{{\rm{Pr}}\left(|\delta_a^+\rangle_{A'_j} |\delta_b^-\rangle_{B'_j}\right)+{\rm{Pr}}\left(|\delta_a^-\rangle_{A'_j}|\delta_b^+\rangle_{B'_j} \right)}{\Pr(\delta_a\delta_b)}\\
=&\frac{p_1^4 p_{1|2\mu}^2}{4 \Tr \rho_{X_1}^{(i,j)}}  \Tr \left\{ \hat{M}_E \bm{\mathcal{P}} \left[  \frac{ |01\rangle_{a_i a_j}+ e^{-{\rm{i}}\delta_a} |10\rangle_{a_ia_j}}{\sqrt{2}} \otimes \frac{|01\rangle_{b_i b_j}- e^{-{\rm{i}}\delta_b} |10\rangle_{b_ib_j}}{\sqrt{2}} \right] \hat{M}^\dagger_E \right\} \\
+&\frac{p_1^4 p_{1|2\mu}^2}{4 \Tr \rho_{X_1}^{(i,j)}}  \Tr \left\{ \hat{M}_E \bm{\mathcal{P}} \left[  \frac{ |01\rangle_{a_i a_j}- e^{-{\rm{i}}\delta_a} |10\rangle_{a_ia_j}}{\sqrt{2}} \otimes \frac{|01\rangle_{b_i b_j}+ e^{-{\rm{i}}\delta_b} |10\rangle_{b_ib_j}}{\sqrt{2}} \right] \hat{M}^\dagger_E \right\}. \\
\end{aligned}
\end{equation}
And the single-photon bit error rate of X-pair preparation under the condition Alice and Bob obtained the results $\delta_a$ and $\delta_b$ ($\delta_{a(b)}\in[0,\pi)$, $|\delta_a-\delta_b|\geq\pi-\Delta$) is
\begin{equation}
\begin{aligned}
e_{X_1}^{\rm{bit}}(\delta_a,\delta_b)=& \frac{{\rm{Pr}}\left(|\delta_a^+\rangle_{A'_j} |\delta_b^+\rangle_{B'_j}\right)+{\rm{Pr}}\left(|\delta_a^-\rangle_{A'_j}|\delta_b^-\rangle_{B'_j} \right)}{\Pr(\delta_a\delta_b)}\\
\end{aligned}
\end{equation}
With the relation $\Tr \rho_{X_1}^{i,j}/ \Tr \tilde\rho_{Z_1}^{i,j}=p_1^2\exp(-2\mu)/p_0^2$ and $e_{Z_1}^{\rm{ph}}$ given in Eq.\eqref{ephz1}, it's easy to verify that 
\begin{equation}
\begin{aligned}
e_{X_1}^{\rm{bit}}(\delta_a,\delta_b)=\left\{
\begin{array}{lc}
e_{Z_1}^{\rm{ph}}(U^{-\delta_a}_{A} U^{-\delta_b}_{B}\tilde\rho_{Z_1}^{i,j} U^{\delta_b}_{B} U^{\delta_a}_{A} ), &|\delta_a-\delta_b|\leq\Delta \vspace{1ex} \\
e_{Z_1}^{\rm{ph}}(U^{-\delta_a}_{A} U^{-(\delta_b+\pi)}_{B} \tilde\rho_{Z_1}^{i,j} U^{(\delta_b+\pi)}_{B} U^{\delta_a}_{A} ), &|\delta_a-\delta_b|\geq\pi-\Delta,\\
\end{array}
\right.
\end{aligned}
\end{equation}
where $U_A^{-\delta_a}|01\rangle_{A_iA_j}=|01\rangle_{A_iA_j}$, $U_A^{-\delta_a}|10\rangle_{A_iA_j}=e^{-{\rm{i}}\delta_a}|10\rangle_{A_iA_j}$ are applied to Alice's local qubits, $U^{-\delta_b}_B$ is similar. This observation confirms that any $e_{X_1}^{\rm{bit}}(\delta_a,\delta_b)$ corresponds to the phase error rate for key bits generated by $\hat{Z}_A\hat{Z}_B$ measurement on the density matrix $U^{-\delta_a}_{A} U^{-\delta_b}_{B}\tilde\rho_{Z_1}^{i,j} U^{\delta_b}_{B} U^{\delta_a}_{A}$ or $U^{-\delta_a}_{A} U^{-(\delta_b+\pi)}_{B} \tilde\rho_{Z_1}^{i,j} U^{(\delta_b+\pi)}_{B} U^{\delta_a}_{A}$. On the other hand, applying $U^{-\delta_a}_{A} U^{-\delta_b}_{B} (U^{-\delta_a}_{A} U^{-(\delta_b+\pi)}_{B})$ or not will not change the distribution of the $\hat{Z}_A\hat{Z}_B$ measurement and potential information leakage to Eve. Hence, without compromising security, the phase error rate $e_{Z_1}^{\rm{ph}}$ can be assumed to be any $e_{X_1}^{\rm{bit}}(\delta_a,\delta_b)$ or its mean value of some domains of $\delta_a,\delta_b$.

Actually, we just calculate $e_{X_1}^{\rm{bit}}(\delta_a,\delta_b)$ satisfying $|\delta_a-\delta_b|\leq\Delta$ or $|\delta_a-\delta_b|\geq\pi-\Delta$ in the $X$-pair preparation. This probability is
\begin{equation}
\begin{aligned}
q_{X_1}=& \int_{0}^{\Delta} \int_{0}^{\delta_a+\Delta} {\rm{Pr}}\left( \delta_a\delta_b \right) 
+ \int_{0}^{\Delta} \int_{\delta_a+\pi-\Delta}^{\pi} {\rm{Pr}}\left( \delta_a\delta_b \right)
+ \int_{\pi-\Delta}^{\pi} \int_{\delta_a-\Delta}^{\pi} {\rm{Pr}}\left( \delta_a\delta_b \right) \\ 
& + \int_{\pi-\Delta}^{\pi} \int_{0}^{\Delta-\pi+\delta_a} {\rm{Pr}}\left( \delta_a\delta_b \right)
+\int_{\Delta}^{\pi-\Delta} \int_{\delta_a-\Delta}^{\delta_a+\Delta} {\rm{Pr}}\left( \delta_a\delta_b \right) \\
=&\frac{2\Delta}{\pi} \Tr \rho_{X_1}^{i,j}.
\end{aligned}
\end{equation}
It should be noted that ${\rm{Pr}}\left( \delta_a\delta_b \right)$ is independent of $\delta_a$ and $\delta_b$. And we come to the main conclusion:  
\begin{equation}
\begin{aligned}
e_{Z_1}^{\rm{ph}}=e_{X_1}^{\rm{bit}}=
&\frac{1}{q_{X_1}} \Bigg[\int_{0}^{\Delta} \int_{0}^{\delta_a+\Delta} {\rm{Pr}}\left( \delta_a\delta_b \right) e_{X_1}^{\rm{bit}}(\delta_a,\delta_b) 
+ \int_{0}^{\Delta}  \int_{\delta_a+\pi-\Delta}^{\pi} {\rm{Pr}}\left( \delta_a\delta_b \right) e_{X_1}^{\rm{bit}}(\delta_a,\delta_b) \\
&  + \int_{\pi-\Delta}^{\pi} \int_{\delta_a-\Delta}^{\pi} {\rm{Pr}}\left( \delta_a\delta_b \right)  e_{X_1}^{\rm{bit}}(\delta_a,\delta_b) 
+ \int_{\pi-\Delta}^{\pi} \int_{0}^{\Delta-\pi+\delta_a} {\rm{Pr}}\left( \delta_a\delta_b \right) e_{X_1}^{\rm{bit}}(\delta_a,\delta_b)\\
&  +\int_{\Delta}^{\pi-\Delta} \int_{\delta_a-\Delta}^{\delta_a+\Delta} {\rm{Pr}}\left( \delta_a\delta_b \right) e_{X_1}^{\rm{bit}}(\delta_a,\delta_b)\Bigg]\\
=& \frac{1}{ 2 \Delta \pi} \Bigg[\int_{0}^{\Delta} \int_{0}^{\delta_a+\Delta} {\rm{d}}\delta_a {\rm{d}}\delta_b  e_{X_1}^{\rm{bit}}(\delta_a,\delta_b) 
+ \int_{0}^{\Delta}  \int_{\delta_a+\pi-\Delta}^{\pi} {\rm{d}}\delta_a {\rm{d}}\delta_b e_{X_1}^{\rm{bit}}(\delta_a,\delta_b) \\
&  + \int_{\pi-\Delta}^{\pi} \int_{\delta_a-\Delta}^{\pi} {\rm{d}}\delta_a {\rm{d}}\delta_b e_{X_1}^{\rm{bit}}(\delta_a,\delta_b) 
+ \int_{\pi-\Delta}^{\pi} \int_{0}^{\Delta-\pi+\delta_a} {\rm{d}}\delta_a {\rm{d}}\delta_b e_{X_1}^{\rm{bit}}(\delta_a,\delta_b)\\
&  +\int_{\Delta}^{\pi-\Delta} \int_{\delta_a-\Delta}^{\delta_a+\Delta}{\rm{d}}\delta_a {\rm{d}}\delta_b e_{X_1}^{\rm{bit}}(\delta_a,\delta_b)\Bigg].\\
\end{aligned}
\end{equation}

In the above proof, we just analyze the relation of $i$-th and $j$-th rounds, we prove that the phase error rate for any individual single-photon $Z$-pair $P^{i,j}$ must be equal to the error rate if single-photon $P^{i,j}$ happens to be $X$-pair. However, it can be expanded to all rounds. It is clear that, for each pairs, this conclusion is satisfied, and the sum of all rounds is also satisfies. In non-asymptotic cases, the phase error rate for the raw key string $\mathbf{Z}_1$ is sampled by the error rate between $\mathbf{X}_1$ and $\mathbf{X}_1'$, where $\mathbf{X}_1$ and $\mathbf{X}_1'$ represent the key strings obtained from single-photon events in $\mathcal{X}$.
To summarize, $e_{X_1}^{\rm{bit}}$ is a random sampling without replacement for $e_{Z_1}^{\rm{ph}}$. 

Then we employ the entropic uncertainty relation for smooth entropies \cite{tomamichel2011uncertainty} to get the length of the secret keys. The auxiliary of Eve after error correction is denoted as $E'$. Due to the quantum leftover hash lemma \cite{renner2008security,tomamichel2011leftover}, a $\varepsilon_{\rm{sec}}$-secret key of length $l_o$ can be extracted from the bit string $\mathbf{Z}$ by applying privacy amplification with two-universal hashing,
\begin{equation}
\begin{aligned}
 2\varepsilon +\frac{1}{2}\sqrt{2^{l_o-H_{\rm{min}}^\varepsilon(\mathbf{Z}|E')}}\leq \varepsilon_{\rm{sec}}, 
\end{aligned}
\end{equation}  
where $H_{\rm{min}}^\varepsilon(\mathbf{Z}|E')$ is the conditional smooth min-entropy, which is employed to quantify the average probability that Eve guesses $\mathbf{Z}$ correctly with $E'$. In the error-correction step, $\lambda_{\rm{EC}}+\log_2 (2/\varepsilon_{\rm{cor}})$ bits are published. By employing a chain-rule inequality for smooth entropies \cite{vitanov2013chain},
\begin{equation}
\begin{aligned}
H_{\rm{min}}^\varepsilon(\mathbf{Z}|E') \geq H_{\rm{min}}^\varepsilon(\mathbf{Z}|E)- \lambda_{\rm{EC}}-\log_2 \frac{2}{\varepsilon_{\rm{cor}}},
\end{aligned}
\end{equation}  
where $E$ is the auxiliary of Eve before error correction. Moreover, $\mathbf{Z}$ can be decomposed into $\mathbf{Z}_1\mathbf{Z}_{\rm{zm}}$, which are the corresponding bit strings due to the single-photon and the other events. We have 
\begin{equation}
\begin{aligned}
H_{\rm{min}}^\varepsilon(\mathbf{Z}|E) \geq H_{\rm{min}}^{\overline{\varepsilon}} (\mathbf{Z}_1|\mathbf{Z}_{\rm{zm}}E)+H_{\rm{min}}^{\varepsilon'}(\mathbf{Z}_{\rm{zm}}|E)-2\log_2\frac{\sqrt{2}}{\hat{\varepsilon}},
\end{aligned}
\end{equation}
where $\varepsilon=2\overline{\varepsilon}+\varepsilon'+\hat{\varepsilon}$ and $H_{\rm{min}}^{\varepsilon'}(\mathbf{Z}_{\rm{zm}}|E) \geq 0$. We can set $\varepsilon'=0$ without compromising security. Here the single-photon component prepared in the $Z$-basis and $X$-basis are mutually unbiased obviously. We employed the string $\mathbf{X}_1(\mathbf{X}_1')$ to denote Alice (Bob) would have obtained if they had measured in the $X$-basis instead of $Z$-basis. By employing the uncertainty relation of smooth min- and max-entropy, we have

\begin{equation}
\begin{aligned}
H_{\rm{min}}^{\overline{\varepsilon}} (\mathbf{Z}_1|\mathbf{Z}_{\rm{zm}}E) \geq n_{Z_1}^{\rm{L}}-H_{\rm{max}}^{\overline{\varepsilon}} (\mathbf{X}_1|\mathbf{X}_1') ,
\end{aligned}
\end{equation} 
where $n_{Z_1}^{\rm{L}}$ is the lower bound of the length of $\mathbf{Z_1}$. And we denote $e_{Z_1}^{\rm{ph}}=(\mathbf{X}_1 \oplus \mathbf{X}_1')/n_{Z_1}^{\rm{L}}$, which is the phase error rate of $n_{Z_1}^{\rm{L}}$ under the $Z$-basis measurement or the bit error rate of $n_{Z_1}^{\rm{L}}$ under the $X$-basis measurement. Actually, $e_{Z_1}^{\rm{ph}}$ cannot directly observe. As is proved in the above, $\langle e_{Z_1}^{\rm{ph}}\rangle=\langle e_{X_1}^{\rm{bit}} \rangle$, $\langle\cdot\rangle$ is employed to denote the expected value. So, $e_{Z_1}^{\rm{ph},U}$ in $n_{Z_1}^{\rm{L}}$ can be bounded by $\langle e_{X_1}^{\rm{bit}} \rangle$ through Chernoff-Bound, where $e_{Z_1}^{{\rm{ph},U}}$ is the estimated value of $e_{Z_1}^{\rm{ph}}$. If the probability that $e_{Z_1}^{\rm{ph}} \geq e_{Z_1}^{{\rm{ph},U}}$ is no larger than $\overline{\varepsilon}^2$,
\begin{equation}
H_{\rm{max}}^{\overline{\varepsilon}} (\mathbf{X}_1|\mathbf{X}_1') \leq n_{Z_1}^{\rm{L}} h(e_{Z_1}^{{\rm{ph},U}}).
\end{equation}

If we choose $\varepsilon_{\rm{sec}}=2(\hat{\varepsilon}+2\overline{\varepsilon})+\varepsilon_{\rm{PA}}$, where $\overline{\varepsilon}=\sqrt{\varepsilon_e+\varepsilon_1}$, $\varepsilon_{1}$ is the probability that the real value of the number of single-photon bits is smaller than $n_{Z_1}^{\rm{L}}$, and $\varepsilon_e$ is the probability that the real value of the phase error rate of single-photon component in $n_{Z_1}^{\rm{L}}$ is bigger than $e_{Z_1}^{\rm{ph},U}$, $\varepsilon_{\rm{PA}}$ is the failure probability of privacy amplification. We have
\begin{equation}
\begin{aligned}
l_o \leq n_{Z_1}^{\rm{L}}\left[ 1-h(e_{Z_1}^{\rm{ph},U})\right]-{\lambda_{\rm{EC}}}-\log_2\frac{2}{\varepsilon_{\rm{cor}}}-2\log_2 \frac{1}{\sqrt{2}\hat{\varepsilon}\varepsilon_{\rm{PA}}}. 
\end{aligned}
\end{equation}

\subsection*{Six-state protocol}

In this section, we analyze the security of the six-state MP-QKD protocol. Similar to the analysis of the original protocol, we omit the decoy state for simplicity. The single-photon component of the joint density matrix of $Z$-pair preparation is the same as the original MP-QKD protocol,
\begin{equation}
\begin{aligned}
\tilde{\rho}_{Z_1}^{i,j}=4 p_0^2 p_1^2 p_{1|\mu}^2 \hat{M}_E \bm{\mathcal{P}}\left[ \frac{|01\rangle_{A_iA_j}|01\rangle_{a_ia_j} + |10\rangle_{A_iA_j}|10\rangle_{a_ia_j}}{\sqrt{2}} \right]\otimes \bm{\mathcal{P}}\left[ \frac{|01\rangle_{B_iB_j}|01\rangle_{b_ib_j} + |10\rangle_{B_iB_j}|10\rangle_{b_ib_j}}{\sqrt{2}} \right] \hat{M}^\dagger_E.
\end{aligned}
\end{equation}
If Alice and Bob take $\hat{X}_A$ and $\hat{X}_B$ measurements, whose eigenstates are 
\begin{equation}
\begin{aligned}
|X_{A}^{+(-)}\rangle_{A_iA_j}:=\frac{ |01\rangle_{A_iA_j} \pm |10\rangle_{A_iA_j}}{\sqrt{2}},
|X_{B}^{+(-)}\rangle_{B_iB_j}:=\frac{ |01\rangle_{B_iB_j} \pm |10\rangle_{B_iB_j}}{\sqrt{2}},
\end{aligned}
\end{equation}
respectively. The single-photon bit error rate of $Z$-pair preparation under these measurements is
\begin{equation}
\begin{aligned}
e_{Z_1}^{X}(\tilde{\rho}_{Z_1}^{i,j})=&
\frac{{\rm{Pr}}(|X_{A}^{+}\rangle_{A_iA_j}|X_{B}^{-}\rangle_{B_iB_j})+
{\rm{Pr}}(|X_{A}^{-}\rangle_{A_iA_j}|X_{B}^{+}\rangle_{B_iB_j})}
{\left[\Pr(|X_{A}^{+}\rangle|X_{B}^{+}\rangle)+
\Pr(|X_{A}^{+}\rangle|X_{B}^{-}\rangle)+
\Pr(|X_{A}^{-}\rangle|X_{B}^{+}\rangle)+
\Pr(|X_{A}^{-}\rangle|X_{B}^{-}\rangle)\right]_{A_iA_j,B_iB_j}}\\
=&\frac{p_0^2p_1^2p_{1|\mu}^2}{\Tr \tilde\rho_{Z_1}^{i,j}}  \Tr \left\{ \hat{M}_E \bm{\mathcal{P}} \left[  \frac{ |01\rangle_{a_i a_j}+|10\rangle_{a_ia_j}}{\sqrt{2}} \otimes \frac{|01\rangle_{b_i b_j}-|10\rangle_{b_ib_j}}{\sqrt{2}} \right] \hat{M}^\dagger_E \right\} \\
+&\frac{p_0^2p_1^2p_{1|\mu}^2}{\Tr \tilde\rho_{Z_1}^{i,j}}  \Tr \left\{ \hat{M}_E \bm{\mathcal{P}} \left[  \frac{ |01\rangle_{a_i a_j}-|10\rangle_{a_ia_j}}{\sqrt{2}} \otimes \frac{|01\rangle_{b_i b_j}+|10\rangle_{b_ib_j}}{\sqrt{2}} \right] \hat{M}^\dagger_E \right\} .\\
\end{aligned}
\end{equation}
And if Alice and Bob take $\hat{Y}_A$ and $\hat{Y}_B$ measurements, whose eigenstates are 
\begin{equation}
\begin{aligned}
|Y_{A}^{+(-)}\rangle_{A_iA_j}:=\frac{ |01\rangle_{A_iA_j} \pm {\rm{i}} |10\rangle_{A_iA_j}}{\sqrt{2}},
|Y_{B}^{+(-)}\rangle_{B_iB_j}:=\frac{ |01\rangle_{B_iB_j} \pm {\rm{i}} |10\rangle_{B_iB_j}}{\sqrt{2}},
\end{aligned}
\end{equation}
respectively. The single-photon bit error rate of $Z$-pair preparation under these measurements is
\begin{equation}
\begin{aligned}
e_{Z_1}^{Y}(\tilde{\rho}_{Z_1}^{i,j})=&
\frac{{\rm{Pr}}(|Y_{A}^{+}\rangle_{A_iA_j}|Y_{B}^{-}\rangle_{B_iB_j})+
{\rm{Pr}}(|Y_{A}^{-}\rangle_{A_iA_j}|Y_{B}^{+}\rangle_{B_iB_j})}
{\left[\Pr(|Y_{A}^{+}\rangle|Y_{B}^{+}\rangle)+
\Pr(|Y_{A}^{+}\rangle|Y_{B}^{-}\rangle)+
\Pr(|Y_{A}^{-}\rangle|Y_{B}^{+}\rangle)+
\Pr(|Y_{A}^{-}\rangle|Y_{B}^{-}\rangle)\right]_{A_iA_j,B_iB_j}}\\
=&\frac{p_0^2p_1^2p_{1|\mu}^2}{\Tr \tilde\rho_{Z_1}^{i,j}}  \Tr \left\{ \hat{M}_E \bm{\mathcal{P}} \left[  \frac{ |01\rangle_{a_i a_j}-{\rm{i}}|10\rangle_{a_ia_j}}{\sqrt{2}} \otimes \frac{|01\rangle_{b_i b_j}+{\rm{i}}|10\rangle_{b_ib_j}}{\sqrt{2}} \right] \hat{M}^\dagger_E \right\} \\
+&\frac{p_0^2p_1^2p_{1|\mu}^2}{\Tr \tilde\rho_{Z_1}^{i,j}}  \Tr \left\{ \hat{M}_E \bm{\mathcal{P}} \left[  \frac{ |01\rangle_{a_i a_j}+{\rm{i}}|10\rangle_{a_ia_j}}{\sqrt{2}} \otimes \frac{|01\rangle_{b_i b_j}-{\rm{i}}|10\rangle_{b_ib_j}}{\sqrt{2}} \right] \hat{M}^\dagger_E \right\} .\\
\end{aligned}
\end{equation}

In the original MP-QKD protocol, if the intensities of $P^{i,j}$ are $\{\mu,\mu\}$, Alice and Bob label the basis as $X$, and they discard the events satisfying $\Delta<|\delta_a-\delta_b|<\pi-\Delta$, where $\delta_a,\delta_b\in [0,\pi)$. In the six-state protocol, when the intensities of $P^{i,j}$ are $\{\mu,\mu\}$, Alice and Bob label the basis of the pairs as $X$ or $Y$ according to the value $r_a$ and $r_b$, $X$-basis if $r_{a(b)}=0$ and $Y$-basis if $r_{a(b)}=1$. And they discard the events satisfying $\Delta<|\delta_a-\delta_b|<\pi/2-\Delta$, where $\delta_a,\delta_b\in [0,\pi/2)$. 

In the $X$- or $Y$-basis preparation of the six-state protocol, the single-photon component of the joint density matrix can be written as
\begin{equation}
\begin{aligned}
\rho_{T_1}^{i,j}=&p_1^4p_{1|2\mu}^2\bm{\mathcal{P}}\left[|11\rangle_{A_iA_j}|11\rangle_{B_iB_j}\right] \\
&\times  \hat{M}_E \bm{\mathcal{P}} \left[ \int_{0}^{\frac{\pi}{2}} \frac{\sqrt{2}{\rm{d}}\delta_a}{\sqrt{\pi}}
\left(\frac{|\delta_a^{+}\rangle|\chi_{\delta_a}^{+}\rangle}{2} 
+\frac{|\delta_a^{+{\rm{i}}}\rangle|\chi_{\delta_a}^{+{\rm{i}}}\rangle}{2} 
+\frac{|\delta_a^{-}\rangle|\chi_{\delta_a}^{-}\rangle}{2} 
+\frac{|\delta_a^{-{\rm{i}}}\rangle|\chi_{\delta_a}^{-{\rm{i}}}\rangle}{2}\right)_{A'_j,a_ia_j}\right]\\
&\otimes  \bm{\mathcal{P}} \left[ \int_{0}^{\frac{\pi}{2}} \frac{\sqrt{2}{\rm{d}}\delta_b}{\sqrt{\pi}}
\left(\frac{|\delta_b^{+}\rangle|\chi_{\delta_b}^{+}\rangle}{2} 
+\frac{|\delta_b^{+{\rm{i}}}\rangle|\chi_{\delta_b}^{+{\rm{i}}}\rangle}{2} 
+\frac{|\delta_b^{-}\rangle|\chi_{\delta_b}^{-}\rangle}{2} 
+\frac{|\delta_b^{-{\rm{i}}}\rangle|\chi_{\delta_b}^{-{\rm{i}}}\rangle}{2}\right)_{B'_j,b_ib_j}\right] \hat{M}_E^\dag,
\end{aligned}
\end{equation} 
where $\delta_a^{+}=\delta_a$, $\delta_a^{+{\rm{i}}}=\delta_a+\pi/2$, $\delta_a^-=\delta_a+\pi$, and $\delta_a^{-{\rm{i}}}=\delta_a+3\pi/2$. Actually, $|\delta_a^+\rangle_{A_j'}$$(|\delta_a^-\rangle_{A_j'})$ corresponds to logical bit 0(1) with $\delta_a$ in the $X$-basis. And $|\delta_a^{+{\rm{i}}}\rangle_{A_j'}$$(|\delta_a^{-{\rm{i}}}\rangle_{A_j'})$ corresponds to logical bit 0(1) with $\delta_a$ in the $Y$-basis. For example, in the six-state protocol, if Alice finds the phase difference between the $i$-th and $j$-th pulse is $3\pi/4$, this is equivalent to that his measurement result of $A'_j$ is $|\delta_a^{+{\rm{i}}}\rangle_{A_j'}$ (logical bit 0 in $Y$-basis) with $\delta_a=\pi/4$ in the entanglement-based protocol. The definitions of $|\delta_b^\pm \rangle_{B_j'}$ and $|\delta_b^{\pm {\rm{i}}} \rangle_{B_j'}$ are similar. And 
\begin{equation}
\begin{aligned}
&|\chi_{\delta_a}^{+(-)}\rangle_{a_ia_j}=\frac{|01\rangle_{a_ia_j} \pm e^{-{\rm{i}}\delta_a} |10\rangle_{a_ia_j}}{\sqrt{2}},
 |\chi_{\delta_a}^{+{\rm{i}}(-{\rm{i}})}\rangle_{a_ia_j}=\frac{|01\rangle_{a_ia_j} \pm {\rm{i}} e^{-{\rm{i}}\delta_a} |10\rangle_{a_ia_j}}{\sqrt{2}},\\
&|\chi_{\delta_b}^{+(-)}\rangle_{b_ib_j}=\frac{|01\rangle_{b_ib_j} \pm e^{-{\rm{i}}\delta_b} |10\rangle_{b_ib_j}}{\sqrt{2}},
 |\chi_{\delta_b}^{+{\rm{i}}(-{\rm{i}})}\rangle_{b_ib_j}=\frac{|01\rangle_{b_ib_j} \pm {\rm{i}} e^{-{\rm{i}}\delta_b} |10\rangle_{b_ib_j}}{\sqrt{2}}.
\end{aligned}
\end{equation}
The probability of Alice and Bob obtaining the measurement results $|\delta_a^+\rangle_{A'_j}$ and $|\delta_b^-\rangle_{B'_j}$ is   
\begin{equation}
\begin{aligned}
\Pr \left( |\delta_a^{+}\rangle_{A'_j}|\delta_b^{-}\rangle_{B'_j} \right)=&
\Tr{\langle \delta_a^{+}|_{A'_j} \langle \delta_b^{-}|_{B'_j} \rho_{T_1}^{i,j} |\delta_a^{+}\rangle_{A'_j}|\delta_b^{-}\rangle_{B'_j}}\\
=& \frac{p_1^4 p_{1|2\mu}^2 {\rm{d}}\delta_a {\rm{d}}\delta_b}{4 \pi^2}  \Tr \left\{ \hat{M}_E \bm{\mathcal{P}} \left[  |\chi_{\delta_a}^{+}\rangle_{a_ia_j}  |\chi_{\delta_b}^{-}\rangle_{b_ib_j} \right] \hat{M}^\dagger_E \right\}.\\
\end{aligned}
\end{equation}
The probabilities of obtaining the other measurement results are similar.

In the key mapping step of the six-state protocol, Bob decides whether to reverse his basis and logical bit according to the relation between $\delta_a$, $\delta_b$ and Charlie's clicks. If $|\delta_a-\delta_b|\leq\Delta$, Bob keeps his basis and logical bit; if $\delta_b-\delta_a \geq \pi/2-\Delta$ and the bases of Alice and Bob are $X$- and $Y$-basis, Bob reverses his basis and his logical bit; if $\delta_b-\delta_a \geq \pi/2-\Delta$ and the bases of two users are $Y$- and $X$-basis, Bob only reverses his basis; if $\delta_a-\delta_b \geq \pi/2-\Delta$ and the bases of two users are $Y$- and $X$-basis, Bob reverses his basis and his logical bit; and if $\delta_a-\delta_b \geq \pi/2-\Delta$ and the bases of two users are $X$- and $Y$-basis, Bob only reverses his basis. In the entanglement-based protocol, Bob implements this operation by changing the classical information of $B'_j$ after the measurement.

After the reverse operation, the yield of the $X$-pair preparation 
\begin{equation}
\begin{aligned}
q_{X_1}=&\int_{0}^{\Delta} \int_{0}^{\delta_a+\Delta} {\rm{Pr}}\left( \delta_a^X\delta_b^X \right) 
+\int_{\Delta}^{\frac{\pi}{2}-\Delta} \int_{\delta_a-\Delta}^{\delta_a+\Delta} {\rm{Pr}}\left( \delta_a^X\delta_b^X \right) 
+ \int_{\frac{\pi}{2}-\Delta}^{\frac{\pi}{2}} \int_{\delta_a-\Delta}^{\frac{\pi}{2}} {\rm{Pr}}\left( \delta_a^X\delta_b^X \right) \\ 
&+ \int_{0}^{\Delta} \int_{\delta_a+\frac{\pi}{2}-\Delta}^{\frac{\pi}{2}} {\rm{Pr}}\left( \delta_a^X\delta_b^Y \right)
+ \int_{\frac{\pi}{2}-\Delta}^{\frac{\pi}{2}} \int_{0}^{\delta_a+\Delta-\frac{\pi}{2}} {\rm{Pr}}\left( \delta_a^X\delta_b^Y \right)\\
=&\frac{\Delta}{\pi} \Tr \rho_{T_1}^{i,j},
\end{aligned}
\end{equation}
where
\begin{equation}
\begin{aligned}
{\rm{Pr}}(\delta_a^X\delta_b^X)=&{\rm{Pr}}\left(|\delta_a^+\rangle_{A'_j}|\delta_b^+\rangle_{B'_j}\right)+
{\rm{Pr}}\left(|\delta_a^+\rangle_{A'_j}|\delta_b^-\rangle_{B'_j}\right)+
{\rm{Pr}}\left(|\delta_a^-\rangle_{A'_j}|\delta_b^+\rangle_{B'_j}\right)+
{\rm{Pr}}\left(|\delta_a^-\rangle_{A'_j}|\delta_b^-\rangle_{B'_j}\right) \\
=&\frac{ {\rm{d}}\delta_a {\rm{d}}\delta_b}{\pi^2} \Tr \rho_{T_1}^{i,j}. \\
{\rm{Pr}}(\delta_a^X\delta_b^Y)=&{\rm{Pr}}\left(|\delta_a^{+}\rangle_{A'_j}|\delta_b^{+{\rm{i}}}\rangle_{B'_j}\right)+
{\rm{Pr}}\left(|\delta_a^{+}\rangle_{A'_j}|\delta_b^{-{\rm{i}}}\rangle_{B'_j}\right)+
{\rm{Pr}}\left(|\delta_a^{-}\rangle_{A'_j}|\delta_b^{+{\rm{i}}}\rangle_{B'_j}\right)+
{\rm{Pr}}\left(|\delta_a^{-}\rangle_{A'_j}|\delta_b^{-{\rm{i}}}\rangle_{B'_j}\right) \\  
=&\frac{ {\rm{d}}\delta_a {\rm{d}}\delta_b}{\pi^2} \Tr \rho_{T_1}^{i,j}.
\end{aligned}
\end{equation}
And the single-photon bit error rate of $X$-pair preparation is
\begin{equation}
\begin{aligned}
e_{X_1}^{\rm{bit}}=&\frac{1}{q_{X_1}}\left\{ \int_{0}^{\Delta} \int_{0}^{\delta_a+\Delta} \left[\Pr \left( |\delta_a^{+}\rangle_{A'_j}|\delta_b^{-}\rangle_{B'_j} \right) + \Pr \left( |\delta_a^{-}\rangle_{A'_j}|\delta_b^{+}\rangle_{B'_j} \right) \right] \right.\\
&+\int_{\Delta}^{\frac{\pi}{2}-\Delta} \int_{\delta_a-\Delta}^{\delta_a+\Delta} \left[\Pr \left( |\delta_a^{+}\rangle_{A'_j}|\delta_b^{-}\rangle_{B'_j} \right) + \Pr \left( |\delta_a^{-}\rangle_{A'_j}|\delta_b^{+}\rangle_{B'_j} \right) \right] \\
&+ \int_{\frac{\pi}{2}-\Delta}^{\frac{\pi}{2}} \int_{\delta_a-\Delta}^{\frac{\pi}{2}} \left[\Pr \left( |\delta_a^{+}\rangle_{A'_j}|\delta_b^{-}\rangle_{B'_j} \right) + \Pr \left( |\delta_a^{-}\rangle_{A'_j}|\delta_b^{+}\rangle_{B'_j} \right) \right] \\
&+ \int_{0}^{\Delta} \int_{\delta_a+\frac{\pi}{2}-\Delta}^{\frac{\pi}{2}} \left[\Pr \left( |\delta_a^{+}\rangle_{A'_j}|\delta_b^{+{\rm{i}}}\rangle_{B'_j} \right) + \Pr \left( |\delta_a^{-}\rangle_{A'_j}|\delta_b^{-{\rm{i}}}\rangle_{B'_j} \right) \right] \\
&\left. + \int_{\frac{\pi}{2}-\Delta}^{\frac{\pi}{2}} \int_{0}^{\delta_a+\Delta-\frac{\pi}{2}} \left[\Pr \left( |\delta_a^{+}\rangle_{A'_j}|\delta_b^{-{\rm{i}}}\rangle_{B'_j} \right) + \Pr \left( |\delta_a^{-}\rangle_{A'_j}|\delta_b^{+{\rm{i}}}\rangle_{B'_j} \right) \right] \right\}. \\
\end{aligned}
\end{equation}
As is proved in the original protocol, it's not restrictive to introduce unitary operations $U^{\delta_a}_A$, $U^{\delta_b}_B$ to the joint density matrix of the $Z$-pair preparation. We come to the main conclusion of the six-state protocol:
\begin{equation}
\begin{aligned}
e_{Z_1}^{X}=&\frac{1}{q_{X_1}}\left\{ \int_{0}^{\Delta} \int_{0}^{\delta_a+\Delta} \Pr \left( \delta_a^X \delta_b^X \right) e_{Z}^{X}\left(U^{-\delta_a}_{A} U^{-\delta_b}_{B}\tilde\rho_{Z_1}^{i,j}U^{\delta_b}_{B}U^{\delta_a}_{A} \right)       \right. + \int_{\Delta}^{\frac{\pi}{2}-\Delta} \int_{\delta_a-\Delta}^{\delta_a+\Delta} \Pr \left( \delta_a^X \delta_b^X \right) e_{Z}^{X}\left(U^{-\delta_a}_{A} U^{-\delta_b}_{B}\tilde\rho_{Z_1}^{i,j} U^{\delta_b}_{B}U^{\delta_a}_{A}\right) \\
&+ \int_{\frac{\pi}{2}-\Delta}^{\frac{\pi}{2}} \int_{\delta_a-\Delta}^{\frac{\pi}{2}} \Pr \left( \delta_a^X \delta_b^X \right) e_{Z}^{X}\left(U^{-\delta_a}_{A} U^{-\delta_b}_{B}\tilde\rho_{Z_1}^{i,j}U^{\delta_b}_{B}U^{\delta_a}_{A} \right) + \int_{0}^{\Delta} \int_{\delta_a+\frac{\pi}{2}-\Delta}^{\frac{\pi}{2}} \Pr \left( \delta_a^X \delta_b^Y \right) e_{Z}^{X}\left(U^{-\delta_a}_{A} U^{-(\delta_b-\frac{\pi}{2})}_{B} \tilde\rho_{Z_1}^{i,j} U^{(\delta_b-\frac{\pi}{2})}_{B} U^{\delta_a}_{A}\right) \\ 
&\left.+  \int_{\frac{\pi}{2}-\Delta}^{\frac{\pi}{2}} \int_{0}^{\delta_a+\Delta-\frac{\pi}{2}} \Pr \left( \delta_a^X \delta_b^Y \right) e_{Z}^{X}\left(U^{-\delta_a}_{A} U^{-(\delta_b+\frac{\pi}{2})}_{B} \tilde\rho_{Z_1}^{i,j} U^{(\delta_b+\frac{\pi}{2})}_{B}U^{\delta_a}_{A} \right) \right\}\\
=&e_{X_1}^{\rm{bit}}.
\end{aligned}
\end{equation}
Similarly,
\begin{equation}
\begin{aligned}
q_{Y_1}=&\int_{0}^{\Delta} \int_{0}^{\delta_a+\Delta} {\rm{Pr}}\left( \delta_a^Y\delta_b^Y \right) 
+\int_{\Delta}^{\frac{\pi}{2}-\Delta} \int_{\delta_a-\Delta}^{\delta_a+\Delta} {\rm{Pr}}\left( \delta_a^Y\delta_b^Y \right) 
+ \int_{\frac{\pi}{2}-\Delta}^{\frac{\pi}{2}} \int_{\delta_a-\Delta}^{\frac{\pi}{2}} {\rm{Pr}}\left( \delta_a^Y\delta_b^Y \right) \\ 
&+ \int_{0}^{\Delta} \int_{\delta_a+\pi-\Delta}^{\frac{\pi}{2}} {\rm{Pr}}\left( \delta_a^Y\delta_b^X \right)
+ \int_{\frac{\pi}{2}-\Delta}^{\frac{\pi}{2}} \int_{0}^{\delta_a+\Delta-\frac{\pi}{2}} {\rm{Pr}}\left( \delta_a^Y\delta_b^X \right)\\
=&\frac{\Delta}{\pi} \Tr \rho_{T_1}^{i,j},
\end{aligned}
\end{equation}
and
\begin{equation}
\begin{aligned}
e_{Z_1}^{Y}=&\frac{1}{q_{Y_1}}\left\{ \int_{0}^{\Delta} \int_{0}^{\delta_a+\Delta} \left[\Pr \left( |\delta_a^{+{\rm{i}}}\rangle_{A'_j}|\delta_b^{-{\rm{i}}}\rangle_{B'_j} \right) + \Pr \left( |\delta_a^{-{\rm{i}}}\rangle_{A'_j}|\delta_b^{+{\rm{i}}}\rangle_{B'_j} \right) \right] \right.\\
&+\int_{\Delta}^{\frac{\pi}{2}-\Delta} \int_{\delta_a-\Delta}^{\delta_a+\Delta} \left[\Pr \left( |\delta_a^{+{\rm{i}}}\rangle_{A'_j}|\delta_b^{-{\rm{i}}}\rangle_{B'_j} \right) + \Pr \left( |\delta_a^{-{\rm{i}}}\rangle_{A'_j}|\delta_b^{+{\rm{i}}}\rangle_{B'_j} \right) \right] \\
&+ \int_{\frac{\pi}{2}-\Delta}^{\frac{\pi}{2}} \int_{\delta_a-\Delta}^{\frac{\pi}{2}} \left[\Pr \left( |\delta_A^{+{\rm{i}}}\rangle_{A'_j}|\delta_b^{-{\rm{i}}}\rangle_{B'_j} \right) + \Pr \left( |\delta_a^{-{\rm{i}}}\rangle_{A'_j}|\delta_b^{+{\rm{i}}}\rangle_{B'_j} \right) \right] \\
&+ \int_{0}^{\Delta} \int_{\delta_a+\frac{\pi}{2}-\Delta}^{\frac{\pi}{2}} \left[\Pr \left( |\delta_a^{+{\rm{i}}}\rangle_{A'_j}|\delta_b^{-}\rangle_{B'_j} \right) + \Pr \left( |\delta_a^{-{\rm{i}}}\rangle_{A'_j}|\delta_b^{+}\rangle_{B'_j} \right) \right] \\
&\left. + \int_{\frac{\pi}{2}-\Delta}^{\frac{\pi}{2}} \int_{0}^{\delta_a+\Delta-\frac{\pi}{2}} \left[\Pr \left( |\delta_a^{+{\rm{i}}}\rangle_{A'_j}|\delta_b^{+}\rangle_{B'_j} \right) + \Pr \left( |\delta_a^{-{\rm{i}}}\rangle_{A'_j}|\delta_b^{-}\rangle_{B'_j} \right) \right] \right\} \\
=&\frac{1}{q_{Y_1}}\left\{ \int_{0}^{\Delta} \int_{0}^{\delta_a+\Delta} \Pr \left( \delta_a^Y \delta_b^Y \right) e_{Z}^{Y}\left(U^{-\delta_a}_{A} U^{-\delta_b}_{B}\tilde\rho_{Z_1}^{i,j}U^{\delta_b}_{B}U^{\delta_a}_{A} \right)       \right. \\
&+ \int_{\Delta}^{\frac{\pi}{2}-\Delta} \int_{\delta_a-\Delta}^{\delta_a+\Delta} \Pr \left( \delta_a^Y \delta_b^Y \right) e_{Z}^{Y}\left(U^{-\delta_a}_{A} U^{-\delta_b}_{B}\tilde\rho_{Z_1}^{i,j}U^{\delta_b}_{B}U^{\delta_a}_{A} \right) \\
&+ \int_{\frac{\pi}{2}-\Delta}^{\frac{\pi}{2}} \int_{\delta_a-\Delta}^{\frac{\pi}{2}} \Pr \left( \delta_a^Y \delta_b^Y \right) e_{Z}^{Y}\left(U^{-\delta_a}_{A} U^{-\delta_b}_{B}\tilde\rho_{Z_1}^{i,j}U^{\delta_b}_{B}U^{\delta_a}_{A} \right) \\
&+ \int_{0}^{\Delta} \int_{\delta_a+\frac{\pi}{2}-\Delta}^{\frac{\pi}{2}} \Pr \left( \delta_a^Y \delta_b^X \right) e_{Z}^{Y}\left(U^{-\delta_a}_{A} U^{-(\delta_b-\frac{\pi}{2})}_{B}  \tilde\rho_{Z_1}^{i,j} U^{(\delta_b-\frac{\pi}{2})}_{B} U^{\delta_a}_{A} \right) \\ 
&\left.+  \int_{\frac{\pi}{2}-\Delta}^{\frac{\pi}{2}} \int_{0}^{\delta_a+\Delta-\frac{\pi}{2}} \Pr \left( \delta_a^Y \delta_b^X \right) e_{Z}^{Y}\left(U^{-\delta_a}_{A} U^{-(\delta_b+\frac{\pi}{2})}_{B} \tilde\rho_{Z_1}^{i,j} U^{(\delta_b+\frac{\pi}{2})}_{B} U^{\delta_a}_{A} \right) \right\}\\
=&e_{Y_1}^{\rm{bit}}.
\end{aligned}
\end{equation}

To summarize, $e_{X_1}^{\rm{bit}}$ and $e_{Y_1}^{\rm{bit}}$ is a random sampling without replacement for $e_{Z_1}^{X}$ and $e_{Z_1}^{Y}$.

Just like the analysis of the original protocol, by employing the the quantum leftover hash lemma, the chain rules, and the uncertainty relation of smooth min- and max-entropy, 
\begin{equation}
\begin{aligned}
&2\varepsilon +\frac{1}{2}\sqrt{2^{l_s-H_{\rm{min}}^\varepsilon(\mathbf{Z}|E')}}\leq \varepsilon_{\rm{sec}}, \\
&H_{\rm{min}}^\varepsilon(\mathbf{Z}|E') \geq H_{\rm{min}}^{\overline{\varepsilon}} (\mathbf{Z}_1|\mathbf{Z}_{\rm{zm}}E)-2\log_2\frac{\sqrt{2}}{\hat{\varepsilon}}-{\lambda_{\rm{EC}}}-\log_2 \frac{2}{\varepsilon_{\rm{cor}}},\\
&H_{\rm{min}}^{\overline{\varepsilon}} (\mathbf{Z}_1|\mathbf{Z}_{\rm{zm}}E) \geq n_{Z_1}^{\rm{L}}\left[1-e_{Z_1}^{\rm{bit,U}}\right]\left[1-h\left(\frac{1-\frac{1}{2}\left(e_{Z_1}^{\rm{bit,U}}+\left(e_{X_1}^{\rm{bit}}+e_{Y_1}^{\rm{bit}}\right)^{\rm{U}}\right)}{1-e_{Z_1}^{\rm{bit,U}}}\right)\right],
\end{aligned}
\end{equation}
where $\varepsilon=2\overline{\varepsilon}+\hat{\varepsilon}$ and $\overline{\varepsilon}=\sqrt{1-(1-\varepsilon_1-\varepsilon_e')(1-\varepsilon_1-\varepsilon_e'')}$. The inequation of $H_{\rm{min}}^{\bar{\varepsilon}} (\mathbf{Z}_1|\mathbf{Z}_{\rm{zm}}E)$ is got by the method which is shown in Appendix A of Ref. \cite{wang2021tight}. Then the secret key length can be given by
\begin{equation}
\begin{aligned}
l_s \leq& n_{Z_1}^{\rm{L}}\left[1-e_{Z_1}^{\rm{bit,U}}\right]\left[1-h\left(\frac{1-\frac{1}{2}\left(e_{Z_1}^{\rm{bit,U}}+\left(e_{X_1}^{\rm{bit}}+e_{Y_1}^{\rm{bit}}\right)^{\rm{U}}\right)}{1-e_{Z_1}^{\rm{bit,U}}}\right)\right]\\
&-{\lambda_{\rm{EC}}}-\log_2 \frac{2}{\varepsilon_{\rm{cor}}}-2\log_2{\frac{1}{\sqrt{2}\hat{\varepsilon}\varepsilon_{\rm{PA}}}},
\end{aligned}
\end{equation} 
where $\varepsilon_{\rm{sec}}=2\hat{\varepsilon}+4\overline{\varepsilon}+\varepsilon_{\rm{PA}}$.

\section*{Supplementary Note B: Simulation of observed values} \label{App2_Simulation of observed values}
For simulating the secret key rate without doing an experiment, we employ a theoretical model to simulate the observed values of the experiment, which include the effective detection number and bit error number of different intensities. Without loss of generality, we assume that the properties of Charlie's two detectors are the same. As is defined in the main text, $k^i_a$ and $k^i_b$ are Alice and Bob's preparation intensities of the $i$-th round, $k^i_a\in\{\mu_a,\nu_a,o\}$, $k^i_b\in\{\mu_b,\nu_b,o\}$. $k={(k^i_a+k^j_a,k^i_b+k^j_b)}$ is the group of the intensities in the $i$-th round and $j$-th round, $k_a=k^i_a+k^j_a$, and $k_b=k^i_b+k^j_b$. 

In the $i$-th round, the response probability of only L or R detector can be given by \cite{ma2012alternative,xie2022breaking}
\begin{equation}
\begin{aligned}
q_{k^i_ak^i_b}^{L,\theta^i}&=(1-p_d)e^{-\frac{k^i_a\eta_a+k^i_b\eta_b}{2}}\left[e^{\sqrt{k^i_a\eta_ak^i_b\eta_b} \cos\theta^i} - (1-p_d) e^{-\frac{k^i_a\eta_a+k^i_b\eta_b}{2}}\right],\\
q_{k^i_ak^i_b}^{R,\theta^i}&=(1-p_d)e^{-\frac{k^i_a\eta_a+k^i_b\eta_b}{2}}\left[e^{-\sqrt{k^i_a\eta_ak^i_b\eta_b} \cos\theta^i} - (1-p_d) e^{-\frac{k^i_a\eta_a+k^i_b\eta_b}{2}}\right],\\
\end{aligned}
\end{equation}
where $\theta^i=\theta^i_a-\theta^i_b$ , $p_d$ is the dark count rate, $\eta_a=\eta_d 10^{-\alpha L_a/10}$ and $\eta_b=\eta_d 10^{-\alpha L_b/10}$ are the overall efficiency of Alice and Bob, $\eta_d$ is the detector efficiency, $\alpha$ is the attenuation coefficient of the fiber, and $L_a$, $L_b$ are the distances between Alice, Bob and Charlie. For simplicity, we define $y=(1-p_d)\exp(-(k^i_a\eta_a+k^i_b\eta_b)/2)$ and $\omega=\sqrt{k^i_a\eta_ak^i_b\eta_b}$. The responsive probability can be revised as
\begin{equation}
\begin{aligned}
q_{k^i_ak^i_b}^{L,\theta^i}&=y\left[e^{\omega \cos \theta^i}-y\right],\\
q_{k^i_ak^i_b}^{R,\theta^i}&=y\left[e^{-\omega \cos \theta^i}-y\right],
\end{aligned}
\end{equation}

As is defined in the main text, $C^i$ is the response of the $i$-th round, $C^i=0$ is the invalid event and $C^i=1$ is the effective event. The average response probability during each round, $p$, can be given by a weighted average of conditional probability,
\begin{equation}
\begin{aligned}
p=& {\rm{Pr}} (C^i=1)=\sum_{k^i_ak^i_b} p_{k^i_a} p_{k^i_b} {\rm{Pr}} (C^i=1|k^i_ak^i_b),
\end{aligned}
\end{equation}    
where the conditional probability ${\rm{Pr}}(C^i=1|k^i_ak^i_b)$ can be given by
\begin{equation}
\begin{aligned}
{\rm{Pr}}(C^i=1|k^i_ak^i_b)&= \int_{0}^{2\pi} \frac{{\rm{d}} \theta^i}{2\pi}  \left( q_{k^i_ak^i_b}^{L,\theta^i} + q_{k^i_ak^i_b}^{R,\theta^i}\right)  =2y[I_0(\omega)-y].
\end{aligned}
\end{equation}
$I_0(x)$ represents the zero-order modified Bessel function of the first kind. For a small value of $x$, we can take the first-order approximation $I_0(x) \approx 1+x^2/4$.

Considering the specific grouping strategy in this paper, the expected pair number generated during each round can be given by
\begin{equation}
r_p(p,l)=\left[\frac{1}{p[1-(1-p)^{l}]}+\frac{1}{p} \right],
\end{equation}
where $p$ is the average response probability during each round and $l$ is the maximal pairing interval. The specific calculation is shown in Ref. \cite{zeng2022quantum}.

In the $Z$-pair and '0'-pair, the number of the effective detections can be given by
\begin{equation}
\begin{aligned}
n_{Z}^{k}&=N r_p \sum_{(k^i_a+k^j_a,k^i_b+k^j_b)=k} {\rm{Pr}} (k^i_a k^j_a k^i_b k^j_b|C^i=C^j=1)\\
&= N r_p \sum_{(k^i_a+k^j_a,k^i_b+k^j_b)=k} \frac{{\rm{Pr}}(k^i_a k^i_b)}{{\rm{Pr}}(C^i=1)} {\rm{Pr}}(C^i=1|k^i_ak^i_b) \frac{{\rm{Pr}}(k^j_a k^j_b)}{{\rm{Pr}}(C^j=1)} {\rm{Pr}}(C^j=1|k^j_a k^j_b) \\
&=\frac{N r_p}{p^2} \sum_{(k^i_a+k^j_a,k^i_b+k^j_b)=k} {\rm{Pr}}(k^i_a k^i_bk^j_a k^j_b) {\rm{Pr}}(C^i=1|k^i_a k^i_b)  {\rm{Pr}}(C^j=1|k^j_a k^j_b), 
\end{aligned}
\end{equation}
where $k\in\{(\mu_A,\mu_B),(\mu_A,\nu_B),(\mu_A,0),(\nu_A,\mu_B),(\nu_A,\nu_B),(\nu_A,0),(0,\mu_B),(0,\nu_B),(0,0)\}$ and ${\rm{Pr}}(k^i_a k^i_b k^j_a k^j_b)=p_{k^i_a} p_{k^i_b} p_{k^j_a} p_{k^j_b}$. Then we consider the error effective detections. For $k\in\{(\mu_A,0),(\nu_A,0),(0,\mu_B),(0,\nu_B),(0,0)\}$, the number of the error effective detections without considering $e_d^Z$ is $m_{Z}^{k,0}=n_{Z}^{k}/2$, where $e_d^Z$ is the misalignment-error of the $Z$-pair. And for $k\in\{(\mu_A,\mu_B),(\mu_A,\nu_B),(\nu_A,\mu_B),(\nu_A,\nu_B)\}$, the number of the error effective detections without considering $e_d^Z$ is
\begin{equation}
\begin{aligned}
m_{Z}^{k,0}&=N r_p \left[{\rm{Pr}} (k^i_a=k^i_b=0|C^i=C^j=1) + {\rm{Pr}} (k^j_a=k^j_b=0|C^i=C^j=1)\right]\\
&=\frac{N r_p}{p^2} \sum_{\substack{k^i_a=k^i_b=0,\\(k^i_a+k^j_a,k^i_b+k^j_b)=k}} {\rm{Pr}}(k^i_a k^i_b k^j_a k^j_b) {\rm{Pr}}(C^i=1|k^i_a k^i_b)  {\rm{Pr}}(C^j=1|k^j_a k^j_b) \\
&+\frac{N r_p}{p^2} \sum_{\substack{k^j_a=k^j_b=0,\\(k^i_a+k^j_a,k^i_b+k^j_b)=k}} {\rm{Pr}}(k^i_a k^i_b k^j_a k^j_b) {\rm{Pr}}(C^i=1|k^i_a k^i_b)  {\rm{Pr}}(C^j=1|k^j_a k^j_b).
\end{aligned}
\end{equation}
And when we take $e_d^Z$ into consideration,
\begin{equation}
\begin{aligned}
m_{Z}^{k}&=(1-e_d^Z) m_{Z}^{k,0} + e_d^Z (n_{Z}^{k}-m_{Z}^{k,0}) .
\end{aligned}
\end{equation}

In the $X$-pair of the original protocol, the number of the effective detections before phase postselection can be given by
\begin{equation}
\begin{aligned}
n_{X}^{k,all}&=\frac{N r_p}{p^2} {\rm{Pr}}(k^i_a k^i_b k^j_a k^j_b) {\rm{Pr}}(C^i=1|k^i_a k^i_b)  {\rm{Pr}}(C^j=1|k^j_a k^j_b),
\end{aligned}
\end{equation}
where $k\in\{(2\mu_A,2\mu_B),(2\mu_A,2\nu_B),(2\mu_A,0),(2\nu_A,2\mu_B),(2\nu_A,2\nu_B),(2\nu_A,0),(0,2\mu_B),(0,2\nu_B)\}$.

In the key mapping step, Alice and Bob take the phase postselection to $\delta_a$ and $\delta_b$. For the events that $k_a=0$ or $k_b=0$ ($k\in\{(2\mu_A,0),(2\nu_A,0),(0,2\mu_B),(0,2\nu_B)\}$), they retain all the data pairs, the number of the reversed effective detections is $n_{X}^{k}=n_{X}^{k,all}$, $m_{X}^{k,0}=n_{X}^{k,all}/2$. And for the other events, the number of the reversed effective detections is
\begin{equation}
\begin{aligned}
n_{X}^{k}=&n_{X}^{k,all}
\frac{ \int_{0}^{2\pi} {\rm{d}}\theta^i \int_{-\Delta}^{\Delta} {\rm{d}}\delta \left( q_{k^i_a k^i_b}^{L,\theta^i} + q_{k^i_a k^i_b}^{R,\theta^i}\right)  \left( q_{k^j_a k^j_b}^{L,\theta^i+\delta} + q_{k^j_a k^j_b}^{R,\theta^i+\delta}\right) }
{ \int_{0}^{2\pi} {\rm{d}} \theta^i \int_{0}^{2\pi} {\rm{d}} \delta \left( q_{k^i_a k^i_b}^{L,\theta^i} + q_{k^i_a k^i_b}^{R,\theta^i}\right) \left( q_{k^j_a k^j_b}^{L,\theta^i+\delta} + q_{k^j_a k^j_b}^{R,\theta^i+\delta}\right)}\\
&+n_{X}^{k,all}\frac{ \int_{0}^{2\pi} {\rm{d}}\theta^i \int_{\pi-\Delta}^{\pi+\Delta} {\rm{d}}\delta \left( q_{k^i_a k^i_b}^{L,\theta^i} + q_{k^i_a k^i_b}^{R,\theta^i}\right)  \left( q_{k^j_a k^j_b}^{L,\theta^i+\delta} + q_{k^j_a k^j_b}^{R,\theta^i+\delta}\right)}
{ \int_{0}^{2\pi} {\rm{d}} \theta^i \int_{0}^{2\pi} {\rm{d}} \delta \left( q_{k^i_a k^i_b}^{L,\theta^i} + q_{k^i_a k^i_b}^{R,\theta^i}\right) \left( q_{k^j_a k^j_b}^{L,\theta^i+\delta} + q_{k^j_a k^j_b}^{R,\theta^i+\delta}\right)}\\
=& \frac{ N r_p}{p^2} \frac{2\Delta}{\pi} {\rm{Pr}}(k^i_a k^i_b k^j_a k^j_b) \\
& \times \left[ y^2 \int_{0}^{2\pi} \frac{{\rm{d}}\theta^i}{2\pi} \int_{-\Delta}^{\Delta} \frac{{\rm{d}}\delta}{2\Delta}  \left(e^{\omega \cos \theta^i} + e^{-\omega \cos \theta^i}\right) \left(e^{\omega \cos(\theta^i+\delta)} + e^{-\omega \cos (\theta^i+\delta)}\right) -8y^3 I_0(\omega)+4y^4\right],
\end{aligned}
\end{equation}
where $\theta^i=\theta^i_a-\theta^i_b$, $\delta=\delta_a-\delta_b$. 
The quantity without considering $e_d^X$ can be given by
\begin{equation}
\begin{aligned}
m_{X}^{k,0}&=\frac{N r_p}{p^2} \frac{2\Delta}{\pi}  {\rm{Pr}}(k^i_a k^i_b k^j_a k^j_b) \int_{0}^{2\pi} \frac{{\rm{d}}\theta^i}{2\pi} \int_{-\Delta}^{\Delta} \frac{{\rm{d}}\delta}{2\Delta} (q_{k^j_a k^j_b}^{L,\theta^i} q_{k^j_a k^j_b}^{R,\theta^i+\delta}+ q_{k^j_a k^j_b}^{R,\theta^i} q_{k^j_a k^j_b}^{L,\theta^i+\delta})   \\
&=\frac{N r_p}{p^2} \frac{2\Delta}{\pi} {\rm{Pr}}(k^i_a k^i_b k^j_a k^j_b) \left[ y^2 \int_{0}^{2\pi}\frac{{\rm{d}}\theta^i}{2\pi}  \int_{-\Delta}^{\Delta}\frac{{\rm{d}}\delta}{2\Delta}  \left(e^{\omega \left(\cos \theta^i- \cos (\theta^i+\delta)\right)}+e^{-\omega \left(\cos \theta^i- \cos (\theta^i+\delta)\right)} \right)   -4y^3 I_0(\omega)+2y^4   \right].
\end{aligned}
\end{equation}
When we take $e_d^X$ into consideration,
\begin{equation}
\begin{aligned}
m_{X}^{k}&=(1-e_d^X) m_{X}^{k,0} + e_d^X (n_{X}^{k}-m_{X}^{k,0}) . 
\end{aligned}
\end{equation}

The simulations of the $X$- and $Y$-pair of the six-state protocol are in the same way. However, in the six-state protocol, we divided the original protocol's $X$-pair into two pairs, the number of the reversed error effective detections of $X$-pair ($Y$-pair), $\overline{m}_{X(Y)}^{k}$, are half of $m_{X}^{k}$.


\section*{Supplementary Note C: Simulation formulas} \label{App3_Simulation formulas}

In this section, we show the specific formulas which are employed to get the secret key length. In the $i$-th round, Alice and Bob prepare weak coherent states $|\sqrt{k^i_a}\exp({\rm{i}}\theta^i_a)\rangle$ and $|\sqrt{k^i_b}\exp({\rm{i}}\theta^i_b)\rangle$. After the mode pairing, basis sifting and key mapping steps, the basis, the key bit and the alignment angle of different pairs are determined. As is proof in the Ref. \cite{zeng2022quantum}, the traditional decoy-state formulas for MDI-QKD can be employed directly by introducing the 'gain' and 'yield'. It should be noted that the 'gain' and 'yield' are analogous values to the MDI's gain and yield, but not the real gain and yield of MP-QKD. In this study, we analyze the secret key length by employing joint constraints \cite{yu2015statistical}. 

\subsection*{Original MP-QKD protocol}
The lower bound of the single-photon effective detection number and the upper bound of phase error rate of the single-photon components in the raw key, which are denoted as $n_{Z_1}^{\rm{L}}$ and $e_{Z_1}^{\rm{ph},U}$, is needed for obtaining the secret key length. Actually, the expected values of the 'yield' and the phase error rate of the single-photon components in the raw key satisfy
\begin{equation}
\begin{aligned}
\langle y_{Z_1} \rangle = \langle y_{X_1} \rangle, \langle e_{Z_1}^{\rm{ph}} \rangle=\langle e_{X_1}^{\rm{bit}} \rangle,
\end{aligned}
\end{equation}
where $\langle y_{X_1} \rangle$ is the expected value of the 'yield' of the $X$-pair, and $\langle e_{X_1}^{\rm{bit}} \rangle$ is the expected value of the single-photon bit error rate in the $X$-pair \cite{jiang2021higher}. So, we can estimate $n_{Z_1}^{\rm{L}}$ by employing $\langle y_{Z_1} \rangle$, $e_{Z_1}^{\rm{ph},U}$ by employing $\langle e_{X_1}^{\rm{bit}} \rangle$ with the Chernoff bound 
\begin{equation}
\begin{aligned}
&n_{Z_1}^{\rm{L}}=O^{\rm{L}}({ \sum_{ \substack{k_a \in \{\mu_a,\nu_a\}, \\ k_b \in \{\mu_b,\nu_b\} }} N_Z^k k_a k_b e^{(-k_a-k_b)}\langle y_{Z_1}\rangle^{\rm{L}},\xi_{y}}), \\
&e_{Z_1}^{\rm{ph},U}=O^{\rm{U}}({n_{Z_1}^{\rm{L}} \langle e_{X_1}^{\rm{bit}}\rangle^{\rm{U}},\xi_{e}})/n_{Z_1}^{\rm{L}},
\end{aligned}
\end{equation}
where $N_Z^k=\frac{N}{2} \sum_{k} {\rm{Pr}}(k^i_a k^i_b k^j_a k^j_b)$. It should be noted that $N_Z^k$ is the expected number of pairs with $k$.
$O^{\rm{L}}(E,\xi)$ and $O^{\rm{U}}(E,\xi)$ are defined in Supplementary Note E. For obtaining the bound of expected values $\langle y_{Z_1} \rangle $ and $\langle e_{X_1}^{\rm{bit}} \rangle$, we employ the decoy-state formulas which are shown in Ref. \cite{yu2015statistical,lu2020efficient}. $a_n$, $a'_n$, $b_n$, and $b'_n$ are employed to denote poisson distribution probabilities of intensities $\nu_a$, $\mu_a$, $\nu_b$, and $\mu_b$,
\begin{equation}
\begin{aligned}
&a_n=\frac{\nu_a^n e^{-\nu_a}}{n!}, a_n'=\frac{\mu_a^n e^{-\mu_a}}{n!},b_n=\frac{\nu_b^n e^{-\nu_b}}{n!}, b_n'=\frac{\mu_b^n e^{-\mu_b}}{n!},
\end{aligned}
\end{equation}
respectively.
If $\frac{a_1'b_2'}{a_1b_2} \leq \frac{a_2'b_1'}{a_2b_1}$, the lower bound of $\langle y_{Z_1}\rangle$ can be estimated by
\begin{equation}
\begin{aligned}
\langle y_{Z_1}\rangle^{\rm{L}}=\frac{Y_+^{\rm{L}}-Y_-^{\rm{U}}}{a_1 a_1' \tilde{b}_{12}},
\end{aligned}
\end{equation}
where $\tilde{b}_{12}=b_1 b_2'-b_1' b_2$, and
\begin{equation}
\begin{aligned}
Y_+^{\rm{L}}&= \rm{min}:\frac{a_1' b_2'}{N_Z^{(\nu,\nu)}} \langle n_Z^{(\nu,\nu)} \rangle
+\frac{a_1 b_2 a_0'}{N_Z^{(0,\mu)}} \langle n_Z^{(0,\mu)} \rangle
+\frac{a_1 b_2 b_0'}{N_Z^{(\mu,0)}} \langle n_Z^{(\mu,0)} \rangle
+\frac{a_1' b_2' a_0 b_0-a_1 b_2 a_0' b_0'}{N_Z^{(0,0)}} \langle n_Z^{(0,0)} \rangle,\\
Y_-^{\rm{U}}&= \rm{max}: \frac{a_1 b_2}{N_Z^{(\mu,\mu)}} \langle n_Z^{(\mu,\mu)} \rangle
+\frac{a_1' b_2' a_0}{N_Z^{(0,\nu)}} \langle n_Z^{(0,\nu)} \rangle
+\frac{a_1' b_2' b_0}{N_Z^{(\nu,0)}} \langle n_Z^{(\nu,0)} \rangle.
\end{aligned}
\end{equation}
Here we employ the method of joint study to obtain $Y_+^{\rm{L}}$ and $Y_-^{\rm{U}}$, which is proposed in Ref. \cite{yu2015statistical}. The core of this method is to use joint constraints between different measured values to limit the influence of statistical fluctuations. For example, $Y_+^{\rm{L}}$ can be simply written as $\rm{min}:\gamma_1g_1+\gamma_2g_2+\gamma_3g_3+\gamma_4g_4$, where $\gamma_i(i\in\{1,2,3,4\})$ is the coefficient behind the expected values, and $g_i(i\in\{1,2,3,4\})$ is the expected value of measured values. In the method of joint study, we not only employ the Chernoff Bound to estimate the lower bound of $g_i$, but also employ $g_i+g_{i'} (i \neq i', i,i'\in\{1,2,3,4\})$, $g_i+g_{i'}+g_{i''} (i \neq i' \neq i'', i,i',i''\in\{1,2,3,4\})$ and $g_1+g_2+g_3+g_4$ to bound the lower bound of $Y_+^{\rm{L}}$. Here different $g_i$ are seen as a same event. The specific descriptions are shown in Supplementary Note D. For simplicity, $Y_+^{\rm{L}}$ and $Y_-^{\rm{U}}$ are rewrote as
\begin{equation}
\begin{aligned}
Y_+^{\rm{L}}&={F}^{\rm{L}} \left(\frac{a_1' b_2'}{N_Z^{(\nu,\nu)}},\frac{a_1 b_2 a_0'}{N_Z^{(0,\mu)}},\frac{a_1 b_2 b_0'}{N_Z^{(\mu,0)}},\frac{a_1' b_2' a_0 b_0-a_1 b_2 a_0' b_0'}{N_Z^{(0,0)}},n_Z^{(\nu,\nu)},n_Z^{(0,\mu)},n_Z^{(\mu,0)},n_Z^{(0,0)},\xi_{y_1},\xi_{y_2},\xi_{y_3},\xi_{y_4}\right),\\
Y_-^{\rm{U}}&={F}^{\rm{U}} \left(\frac{a_1 b_2}{N_Z^{(\mu,\mu)}},\frac{a_1' b_2' a_0}{N_Z^{(0,\nu)}},\frac{a_1' b_2' b_0}{N_Z^{(\nu,0)}},0,n_Z^{(\mu,\mu)},n_Z^{(0,\nu)},n_Z^{(\nu,0)},0,\xi_{y_5},\xi_{y_6},\xi_{y_7},0\right).
\end{aligned}
\end{equation}
The functions ${F}^{\rm{L}}(\cdot)$ and ${F}^{\rm{U}}(\cdot)$ are analyzed by employing joint constraints with the Chernoff bound, which is shown in Supplementary Note D. In the case of $\frac{a_1'b_2'}{a_1b_2} > \frac{a_2'b_1'}{a_2b_1}$, the lower bound of $\langle y_{Z_1} \rangle$ can be obtained by making the exchange between $a_n$ and $b_n$, and the exchange between $a_n'$ and $b_n'$, for $n=1,2$. 

Moreover, the upper bound of $\langle e_{X_1}^{\rm{bit}} \rangle$ satisfies
\begin{equation}
\begin{aligned}
\langle e_{X_1}^{\rm{bit}} \rangle^{\rm{U}}=\frac{T_+^{\rm{U}}-T_-^{\rm{L}}}{a_1 b_1 \langle y_{Z_1}\rangle^{\rm{L}}},
\end{aligned}
\end{equation}
where 
\begin{equation}
\begin{aligned}
T_+^{\rm{U}}=&{F}^{\rm{L}} \left(\frac{1}{N_X^{(2\nu,2\nu)}},\frac{a_0 b_0}{N_X^{(0,0)}},0,0,\right.\left.m_X^{(2\nu,2\nu)},m_X^{(0,0)},0,0,\xi_{e_1},\xi_{e_2},0,0\right),\\
T_-^{\rm{L}}=&{F}^{\rm{U}} \left(\frac{a_0}{N_X^{(0,2\nu)}},\frac{b_0}{N_X^{(2\nu,0)}},0,0,\right.\left.m_X^{(0,2\nu)},m_X^{(2\nu,0)},0,0,\xi_{e_3},\xi_{e_4},0,0\right).
\end{aligned}
\end{equation}
Here $N_X^{(2\nu,2\nu)}=\frac{N\Delta}{\pi} p_\nu^4 $, $N_X^{(0,2\nu)}=N_X^{(2\nu,0)}=\frac{N}{2} p_\nu^2 p_o^2$, and $N_X^{(0,0)}=\frac{N}{2} p_o^4$.

In the universally composable framework, $\varepsilon_1$ is the probability that the real value of the number of single-photon bits is smaller than $n_{Z_1}^{\rm{L}}$, and $\varepsilon_e$ is the probability that the real value of the phase error rate of single-photon component in $n_Z^{\rm{L}}$ is bigger than $e_{Z_1}^{\rm{ph},U}$. With the calculation method above, the failure probability of the estimation of $n_{Z_1}^{\rm{L}}$ and $e_{Z_1}^{\rm{ph},U}$ are
\begin{equation}
\begin{aligned}
\varepsilon_1=&\xi_y+\xi_{y_1}+\xi_{y_2}+\xi_{y_3}+\xi_{y_4}+\xi_{y_5}+\xi_{y_6}+\xi_{y_7},\\
\varepsilon_e=&\xi_e+\xi_{e_1}+\xi_{e_2}+\xi_{e_3}+\xi_{e_4}.
\end{aligned}
\end{equation}  

\subsection*{Six-state MP-QKD protocol}
For the six-state MP-QKD protocol, we need $n_{Z_1}^{\rm{L}}$, $e_{Z_1}^{\rm{bit,U}}$, and $(e_{X_1}^{\rm{bit}}+e_{Y_1}^{\rm{bit}})^{\rm{U}}$ to obtain the secret key length. $e_{Z_1}^{\rm{bit,U}}$ is the upper bound of the bit error rate in the single-photon component of the raw key. $(e_{X_1}^{\rm{bit}}+e_{Y_1}^{\rm{bit}})^{\rm{U}}$ is the upper bound of the sum of the bit error rate if Alice and Bob measure the single-photon component of the raw key in $\hat{X}$ and $\hat{Y}$.
 
The upper bound of $\langle e_{Z_1}^{\rm{bit}} \rangle$ satisfies
\begin{equation}
\begin{aligned}
\langle e_{Z_1}^{\rm{bit}} \rangle^{\rm{U}}=\frac{T_+^{'\rm{U}}-T_-^{'\rm{L}}}{a_1 b_1 \langle y_{Z_1}\rangle^{\rm{L}}},
\end{aligned}
\end{equation}
where 
\begin{equation}
\begin{aligned}
T_+^{'\rm{U}}=&{F}^{\rm{L}} \left(\frac{1}{N_Z^{(\nu,\nu)}},\frac{a_0 b_0}{N_Z^{(0,0)}},0,0,\right.\left.m_Z^{(\nu,\nu)},m_Z^{(0,0)},0,0,\xi'_{e_1},\xi'_{e_2},0,0\right),\\
T_-^{'\rm{L}}=&{F}^{\rm{U}} \left(\frac{a_0}{N_Z^{(0,2\nu)}},\frac{b_0}{N_Z^{(2\nu,0)}},0,0,\right.\left.m_Z^{(0,\nu)},m_Z^{(\nu,0)},0,0,\xi'_{e_3},\xi'_{e_4},0,0\right).
\end{aligned}
\end{equation}

The upper bound of $\langle e_{X_1}^{\rm{bit}}+e_{Y_1}^{\rm{bit}}\rangle$ is
\begin{equation}
\begin{aligned}
\langle e_{X_1}^{\rm{bit}}+e_{Y_1}^{\rm{bit}} \rangle^{\rm{U}}=\frac{T_+^{'\rm{U}}-T_-^{'\rm{L}}}{a_1 b_1 \langle y_{Z_1}\rangle^{\rm{L}}},
\end{aligned}
\end{equation}
where 
\begin{equation}
\begin{aligned}
T_+^{'\rm{U}}&={F}^{\rm{L}} \left(\frac{1}{\overline{N}_{X}^{(2\nu,2\nu)}},\frac{a_0 b_0}{\overline{N}_{X}^{(0,0)}},0,0,\right.\left.\overline{m}_X^{(2\nu,2\nu)}+\overline{m}_Y^{(2\nu,2\nu)},\overline{m}_X^{(0,0)}+\overline{m}_Y^{(0,0)},0,0,\xi''_{e_1},\xi''_{e_2},0,0\right),\\
T_-^{'\rm{L}}&={F}^{\rm{U}} \left(\frac{a_0}{\overline{N}_{X}^{(0,2\nu)}},\frac{b_0}{\overline{N}_{X}^{(2\nu,0)}},0,0,\right.\left.\overline{m}_X^{(0,2\nu)}+\overline{m}_Y^{(0,2\nu)},\overline{m}_X^{(2\nu,0)}+\overline{m}_Y^{(0,2\nu)},0,0,\xi''_{e_3},\xi''_{e_4},0,0\right),
\end{aligned}
\end{equation}
Here $\overline{N}_X^{k}=\frac{N_X^{k}}{2} \left(k\in\{(2\nu,2\nu),(2\nu,0),(0,2\nu),(0,0)\}\right)$.

Then we can get $e_{Z_1}^{\rm{bit,U}}$, and $(e_{X_1}^{\rm{bit}}+e_{Y_1}^{\rm{bit}})^{\rm{U}}$ by Chernoff bound,
\begin{equation}
\begin{aligned}
&e_{Z_1}^{\rm{bit,U}}=O^{\rm{U}}({n_{Z_1}^{\rm{L}} \langle e_{Z_1}^{\rm{bit}}\rangle^{\rm{U}},\xi'_{e}})/n_{Z_1}^{\rm{L}},\\
&(e_{X_1}^{\rm{bit}}+e_{Y_1}^{\rm{bit}})^{\rm{U}}=O^{\rm{U}}({n_{Z_1}^{\rm{L}} \langle e_{X_1}^{\rm{bit}}+e_{Y_1}^{\rm{bit}} \rangle^{\rm{U}},\xi''_{e}})/n_{Z_1}^{\rm{L}}.\\
\end{aligned}
\end{equation}

In the universally composable framework, $\varepsilon'_e$ is the probability that the real value of the bit error rate of single-photon component in $n_{Z_1}^{\rm{L}}$ under $Z$-pair measurement is bigger than $e_{Z_1}^{\rm{bit,U}}$. $\varepsilon''_e$ is the probability that if Alice and Bob take $X$-pair measurement and $Y$-pair measurement to $n_{Z_1}^{\rm{L}}$, the real value of the sum of bit error rate is bigger than $\left(e_{X_1}^{\rm{bit}}+e_{Y_1}^{\rm{bit}}\right)^{\rm{U}}$. With the calculation method above, the failure probability of the estimation of $e_{Z_1}^{\rm{bit,U}}$ and $\left(e_{X_1}^{\rm{bit}}+e_{Y_1}^{\rm{bit}}\right)^{\rm{U}}$ are
\begin{equation}
\begin{aligned}
\varepsilon'_e=&\xi'_e+\xi'_{e_1}+\xi'_{e_2}+\xi'_{e_3}+\xi'_{e_4},\\
\varepsilon''_e=&\xi''_e+\xi''_{e_1}+\xi''_{e_2}+\xi''_{e_3}+\xi''_{e_4}.
\end{aligned}
\end{equation}

\section*{Supplementary Note D: Analytic results of joint constraints} \label{App4_Analytic results of joint constraints}
Here we give the analytic results of joint constraints, which can be write into function ${F}^{\rm{L}}$ and ${F}^{\rm{U}}$. The problem of obtaining the minimum value of function $F$ can be abstracted into 
\begin{equation}
\begin{aligned}
\min\limits_{g_1,g_2,g_3,g_4} & F=\gamma_1g_1+\gamma_2g_2+\gamma_3g_3+\gamma_4g_4, \\
s.t.\quad&g_1 \geq E^{\rm{L}}(\widetilde{g_1},\xi_1),\\
&g_2 \geq E^{\rm{L}}(\widetilde{g_2},\xi_1),\\
&g_3 \geq E^{\rm{L}}(\widetilde{g_3},\xi_1),\\
&g_4 \geq E^{\rm{L}}(\widetilde{g_4},\xi_1),\\
&g_1+g_2 \geq E^{\rm{L}}(\widetilde{g_1}+\widetilde{g_2},\xi_2),\\
&g_1+g_3 \geq E^{\rm{L}}(\widetilde{g_1}+\widetilde{g_3},\xi_2),\\
&g_1+g_4 \geq E^{\rm{L}}(\widetilde{g_1}+\widetilde{g_4},\xi_2),\\
&g_2+g_3 \geq E^{\rm{L}}(\widetilde{g_2}+\widetilde{g_3},\xi_2),\\
&g_2+g_4 \geq E^{\rm{L}}(\widetilde{g_2}+\widetilde{g_4},\xi_2),\\
&g_3+g_4 \geq E^{\rm{L}}(\widetilde{g_3}+\widetilde{g_4},\xi_2),\\
&g_1+g_2+g_3 \geq E^{\rm{L}}(\widetilde{g_1}+\widetilde{g_2}+\widetilde{g_3},\xi_3),\\
&g_1+g_2+g_4 \geq E^{\rm{L}}(\widetilde{g_1}+\widetilde{g_2}+\widetilde{g_4},\xi_3),\\
&g_2+g_3+g_4 \geq E^{\rm{L}}(\widetilde{g_2}+\widetilde{g_3}+\widetilde{g_4},\xi_3),\\
&g_1+g_2+g_3+g_4 \geq E^{\rm{L}}(\widetilde{g_1}+\widetilde{g_2}+\widetilde{g_3}+\widetilde{g_4},\xi_3),
\end{aligned}
\end{equation}
where $\gamma_1,\gamma_2,\gamma_3,\gamma_4,g_1,g_2,g_3,g_4,\widetilde{g_1},\widetilde{g_2},\widetilde{g_3},\widetilde{g_4}$ all are positive and $E^{\rm{L}}(O,\xi)$ is Chernoff bound which is presented in Supplementary Note E. In this paper, $\gamma_1$, $\gamma_2$, $\gamma_3$, and $\gamma_4$ are coefficients. $g_1$, $g_2$, $g_3$, and $g_4$ are the expected values of $\widetilde{g_1}$, $\widetilde{g_2}$, $\widetilde{g_3}$, and $\widetilde{g_4}$ which are measured in experiments, respectively.
  
We rearrange $\{\gamma_1,\gamma_2,\gamma_3,\gamma_4\}$ in the ascending order, the new sequence is denoted by $\{\gamma_1',\gamma_2',\gamma_3',\gamma_4'\}$, $\{\widetilde{g_1}',\widetilde{g_2}',\widetilde{g_3}',\widetilde{g_4}'\}$ as the corresponding rearrange of $\{\widetilde{g_1},\widetilde{g_2},\widetilde{g_3},\widetilde{g_4}\}$ according to the ascending order of $\{\gamma_1,\gamma_2,\gamma_3,\gamma_4\}$. In this way, the lower bound of ${F}$ can be wrote as:
\begin{equation}
\begin{aligned}
{F}^{\rm{L}}&(\gamma_1,\gamma_2,\gamma_3,\gamma_4,\widetilde{g_1},\widetilde{g_2},\widetilde{g_3},\widetilde{g_4},\xi_1,\xi_2,\xi_3,\xi_4)\\
=&\gamma_1'E^{\rm{L}}(\widetilde{g_1}'+\widetilde{g_2}'+\widetilde{g_3}'+\widetilde{g_4}',\xi_4)+(\gamma_2'-\gamma_1')E^{\rm{L}}(\widetilde{g_2}'+\widetilde{g_3}'+\widetilde{g_4}',\xi_3)\\
&+(\gamma_3'-\gamma_2')E^{\rm{L}}(\widetilde{g_3}'+\widetilde{g_4}',\xi_2)+(\gamma_4'-\gamma_3')E^{\rm{L}}(\widetilde{g_4}',\xi_1).
\end{aligned}
\end{equation}
And if we want to get the upper bound of ${F}$, we can just replace $E^{\rm{L}}(O,\xi)$ by $E^{\rm{U}}(O,\xi)$:
\begin{equation}
\begin{aligned}
{F}^{\rm{U}}&(\gamma_1,\gamma_2,\gamma_3,\gamma_4,\widetilde{g_1},\widetilde{g_2},\widetilde{g_3},\widetilde{g_4},\xi_1,\xi_2,\xi_3,\xi_4)\\
=&\gamma_1'E^{\rm{U}}(\widetilde{g_1}'+\widetilde{g_2}'+\widetilde{g_3}'+\widetilde{g_4}',\xi_4)+(\gamma_2'-\gamma_1')E^{\rm{U}}(\widetilde{g_2}'+\widetilde{g_3}'+\widetilde{g_4}',\xi_3)\\
&+(\gamma_3'-\gamma_2')E^{\rm{U}}(\widetilde{g_3}'+\widetilde{g_4}',\xi_2)+(\gamma_4'-\gamma_3')E^{\rm{U}}(\widetilde{g_4}',\xi_1).
\end{aligned}
\end{equation}

\section*{Supplementary Note E: Chernoff bound}\label{App5_Chernoff bound}
The Chernoff bound is useful in estimating the expected values from their observed values or estimating the observed values from their expected values. Let $X_1,X_2,...,X_n$ be a set of independent Bernoulli random samples (it can be in different distribution), and let $X=\sum_{i=1}^n X_i$. 
The observed value of $X$ which denotes $O$ is unknown and its expected value $E$ is known. In this situation, we have
\begin{align}
O^{\rm{L}}(E,\xi)=[1-\delta_1(E,\xi)]E,\\
O^{\rm{U}}(E,\xi)=[1+\delta_2(E,\xi)]E,
\end{align} 
where $\delta_1(E,\xi)$ and $\delta_2(E,\xi)$ can be got by solving the following equations:
\begin{align}
\left(\frac{e^{-\delta_1}}{(1-\delta_1)^{1-\delta_1}}\right)^E=\xi,\\
\left(\frac{e^{\delta_2}}{(1+\delta_2)^{1+\delta_2}}\right)^E=\xi,
\end{align}
where $\xi$ is the failure probability.

When the expected value $E$ is unknown and its observed value $O$ is known, we have
\begin{align}
E^{\rm{L}}(O,\xi)=\frac{O}{1+\delta_1'(O,\xi)},\\
E^{\rm{U}}(O,\xi)=\frac{O}{1-\delta_2'(O,\xi)},
\end{align} 
where $\delta_1'(O,\xi)$ and $\delta_2'(O,\xi)$ can be got by solving the following equations:
\begin{align}
\left(\frac{e^{\delta_1'}}{(1+\delta_1')^{1+\delta_1'}}\right)^\frac{O}{1+\delta_1'}=\xi,\\
\left(\frac{e^{-\delta_2'}}{(1-\delta_2')^{1-\delta_2'}}\right)^\frac{O}{1-\delta_2'}=\xi.
\end{align}

\hfill

\noindent {\bf DATA AVAILABILITY}

\noindent The data that support the findings of this study are available from the corresponding author upon reasonable request.

\hfill

\noindent {\bf CODE AVAILABILITY}

\noindent Source codes of the plots are available from the corresponding authors on request.

\hfill

\noindent{\bf ACKNOWLEDGEMENTS}

\noindent This work has been supported by the National Key Research and Development Program of China (Grant No. 2020YFA0309802), the National Natural Science Foundation of China (Grant Nos. 62171424, 61961136004, 62105318, 62271463), China Postdoctoral Science Foundation (2021M693098, 2022M723064), Prospect and Key Core Technology Projects of Jiangsu provincial key R \& D Program (BE2022071), the Fundamental Research Funds for the Central Universities.

\hfill

\noindent{\bf COMPETING INTERESTS}

\noindent The authors declare that they have no competing interests.

\hfill

\noindent{\bf AUTHOR CONTRIBUTIONS}

\noindent Zhen-Qiang Yin, Rong Wang, Ze-Hao Wang conceived the basic idea of the security proof. Zhen-Qiang Yin and Ze-Hao Wang finished the details of the security proof. Ze-Hao Wang performed the numerical simulations. All the authors contributed to discussing the main ideas of the security proof, checking the validity of the results, and writing the paper.

\end{widetext}

\hfill

\noindent{\bf REFERENCES}

\bibliography{sample}

\end{document}